\documentclass[11pt, a4paper]{article}
\pdfoutput=1

\usepackage{amsmath,amssymb}
\usepackage{multirow}
\usepackage[utf8]{inputenc}
\usepackage{bm}
\usepackage{cite}
\usepackage[footnotesize]{caption}

\usepackage{ifpdf}
\ifpdf        
  \usepackage{graphicx, hyperref, xcolor}     
\else     
  \usepackage[dvipdfmx]{graphicx, hyperref, xcolor}     
 \fi
\usepackage{subcaption}

\usepackage{tikz}
\usepackage{tikz-feynman}
\tikzfeynmanset{compat=1.0.0}
 
\definecolor{rossoferrari}{HTML}{D9073D}
\definecolor{mediumblue}{HTML}{0000CD}
\definecolor{forestgreen}{HTML}{228B22}
\definecolor{desy_blue}{HTML}{009EE2}
\definecolor{desy_orange}{HTML}{FD8800}
\hypersetup{
setpagesize=false,
bookmarksnumbered=true,%
bookmarksopen=true,%
colorlinks=true,%
linkcolor=rossoferrari,
urlcolor=mediumblue,
citecolor=mediumblue,
linktocpage=false,
}

\usepackage[height=8.85in,width=6.8in]{geometry}

\usepackage{fourier}

\makeatletter
\@addtoreset{equation}{section}

\makeatother

\newcommand{\abs}[1]{\left\lvert #1 \right\rvert}
\newcommand{\dd}{\mathrm{d}}

\begin{document}

\begin{titlepage}

\begin{center}

\hfill DESY 20-130

\vskip 1.in

{\Huge \bf
Renormalization Group Equations\\ 
of Higgs-$\bm{R^2}$ Inflation\\
}

\vskip .8in

{\Large
Yohei Ema, Kyohei Mukaida, Jorinde van de Vis
}

\vskip .3in

{\em DESY, Notkestra{\ss}e 85, D-22607 Hamburg, Germany}

\end{center}
\vskip .6in

\begin{abstract}
\noindent
We derive one- and two-loop renormalization group equations (RGEs) of 
Higgs-$R^2$ inflation.
This model has a non-minimal coupling between the Higgs and the Ricci scalar 
and a Ricci scalar squared term on top of the standard model.
The RGEs derived in this paper are valid as long as the energy scale of interest (in the Einstein frame)
is below the Planck scale.
We also discuss implications to the inflationary predictions
and the electroweak vacuum metastability.

\end{abstract}

\end{titlepage}

\renewcommand{\thepage}{\arabic{page}}
\setcounter{page}{1}

\tableofcontents
\pagebreak
\renewcommand{\thepage}{\arabic{page}}
\renewcommand{\thefootnote}{$\natural$\arabic{footnote}}
\setcounter{footnote}{0}

\section{Introduction}
\label{sec:intro}

Higgs inflation~\cite{Futamase:1987ua,CervantesCota:1995tz,Bezrukov:2007ep} stands out among many other models of inflation as it is minimal and consistent with the cosmic microwave background (CMB) observation~\cite{Akrami:2018odb}.
It introduces a non-minimal coupling to gravity, 
\begin{align}
	\mathcal{L}_\xi = \xi R \abs{H}^2,
\end{align}
with $H$ the standard model (SM) Higgs doublet, $R$ the Ricci scalar, and $\xi$ the non-minimal coupling,
so that the Higgs potential becomes flat in the Einstein frame for $\abs{H} \gtrsim M_P / \xi$.
The CMB normalization requires $\xi^2 \sim 2 \times 10^9 \lambda$, indicating a large non-minimal coupling, $\xi \gg 1$, unless the Higgs quartic coupling $\lambda$ is tiny.
However, such a large non-minimal coupling causes a strong coupling problem, where the model loses perturbative unitarity at $M_P / \xi$, well below the Planck scale~\cite{Burgess:2009ea,Barbon:2009ya,Burgess:2010zq,Hertzberg:2010dc}.
While this problem does not necessarily ruin the model's validity during inflation~\cite{Bezrukov:2010jz}, during preheating NG bosons (or equivalently longitudinal gauge bosons) with energy exceeding $M_P / \xi$ are produced, threatening perturbative unitarity~\cite{DeCross:2015uza, Ema:2016dny, Sfakianakis:2018lzf}.\footnote{
	The NG boson production itself may be affected 
	by higher dimensional operators~\cite{Hamada:2020kuy},
	but the conclusion that we need knowledge of the UV completion 
	to understand preheating is anyway unchanged.
}
It is thus desirable to push up the scale $M_P/\xi$ where the theory becomes strongly coupled.

Higgs-$R^2$ inflation is a natural solution to the strong coupling problem of Higgs inflation~\cite{Ema:2017rqn,Gorbunov:2018llf}. In addition to the non-minimal couplings term, it also includes the $R^2$-term
\begin{align}
	\mathcal L_\alpha = \alpha R^2.
\end{align}
The $R^2$-term is not only required for the renormalizability of Higgs inflation~\cite{tHooft:1974toh}, but is enhanced for a large $\xi$ because of quantum corrections~\cite{Salvio:2015kka,Calmet:2016fsr,Ema:2017rqn,Ghilencea:2018rqg,Ema:2019fdd,Ema:2020zvg}.
The renormalization group equation (RGE) implies a natural value of the coupling to be $\alpha \sim \xi^2$, which gives rise to a scalaron degree of freedom~\cite{Starobinsky:1980te,Barrow:1983rx,Whitt:1984pd,Barrow:1988xh}
with $m_\sigma \sim M_P / \sqrt{\alpha} \sim M_P / \xi$ right at the strong coupling scale of 
Higgs inflation.
In the same way as the SM Higgs unitarizes the Fermi theory at the electroweak (EW) scale, the scalaron unitarizes  Higgs inflation at $m_\sigma \sim M_P / \xi$, as shown in Refs.~\cite{Ema:2019fdd,Ema:2020zvg}.
Remarkably, the Higgs-$R^2$ model is unitary up to $M_P$ and its quantum corrections never induce other operators (such as the $R^n$-terms with $n \geq 3$) below $M_P$~\cite{Ema:2020zvg}.\footnote{
	This fact is related to the renormalizability of Quadratic Gravity~\cite{Weinberg:1974tw,Deser:1975nv,Stelle:1976gc,Barvinsky:2017zlx,Salvio:2018crh}.
}
Therefore, by computing the beta functions, we can connect the observational input at the EW scale to the UV parameters without ambiguities as long as the scale of our interest is below $M_P$.

In this paper, the beta functions of the Higgs-$R^2$ inflation up to two-loop which are applicable below the Planck scale are presented for the first time.
While the spin-2 part of the metric, \textit{i.e.}, the graviton, only couples to the other fields with Planck-suppressed interactions, the large value of $\xi$ enhances the coupling of matter fields to the conformal mode of the metric $\varphi$ defined by
\begin{align}
		g_{\mu\nu} = e^{2\varphi} \tilde{g}_{\mu\nu},
	\quad
	\mathrm{Det}\left[\tilde{g}_{\mu\nu}\right] = -1.
\end{align}
Hence, as long as we are interested in the beta functions below $M_P$, we may extract the conformal mode of the metric and neglect the contributions from the graviton.
By doing so, we show that the Higgs-$R^2$ model below $M_P$ can be written as a linear sigma model (LSM) of the Higgs, the scalaron, and the conformal mode of the metric with interaction terms which are renormalizable.
This observation dramatically simplifies the calculations. Moreover, the general RGEs up to two-loop for the scalar/gauge/fermion system with renormalizable interactions are already known in the literature~\cite{Machacek:1983tz,Machacek:1983fi,Machacek:1984zw,Luo:2002ti}.
We provide a systematic approach to obtain the RGEs of the Higgs-$R^2$ model from these existing studies.

We then study phenomenological consequences of the obtained RGEs.
Above the scalaron mass scale, the presence of the scalaron gives the threshold correction to the Higgs four-point
interaction~\cite{Ema:2017rqn} and induces the Higgs mass term 
and the cosmological constant of the order of the scalaron mass via the RGEs.
We discuss its implications for the inflationary predictions and the stability of the EW vacuum.

The organization of this paper is as follows.
Sec.~\ref{sec:rge} is the main part of this paper.
In Sec.~\ref{subsec:weyl}, we extract the conformal mode from the rest of the metric. 
In Sec.~\ref{subsec:gauge},
we discuss our gauge fixing condition,
and show that we can ignore the spin-2 part of the gravity, \textit{i.e.}, the graviton,
below the Planck scale.
The full one- and two-loop beta functions of the Higgs-$R^2$ theory 
are then shown in Secs.~\ref{subsec:one_loop} and~\ref{subsec:two_loop},
with computational details given in App.~\ref{app:one_loop} and~\ref{app:two_loop},
respectively.
Phenomenological implications to the inflationary prediction 
and the EW vacuum metastability
are discussed in Secs.~\ref{sec:inf} and~\ref{sec:quartic}, respectively.
Finally Sec.~\ref{sec:conclusion} is devoted to conclusion and discussion.
We follow the convention of Ref.~\cite{Ema:2020zvg} throughout this paper.

\section{Renormalization group equation}
\label{sec:rge}
In this section, we derive the one- and two-loop RGEs of Higgs-$R^2$ inflation
that are valid below the Planck scale.
The action in the Jordan frame is given by
\begin{align}
	S =& \int \dd^4 x \sqrt{-g} 
	\left[\frac{M_P^2}{2}R\left(1+\frac{2\xi \abs{H}^2}{M_P^2}\right) 
	+ g^{\mu\nu}D_\mu H^\dagger D_\nu H
	- m^2\abs{H}^2 - \lambda \abs{H}^4 + \alpha R^2 - \Lambda
	\right]
	+ S_{\psi + A}\left[g_{\mu\nu}\right],
	\label{eq:action_Jordan}
\end{align}
where $g_{\mu\nu}$ is the spacetime metric with $g$ its determinant, $R$ is the Ricci scalar,
$M_P$ is the (reduced) Planck mass, $H$ is the Higgs doublet, $D_\mu$ is its covariant derivative,
$m$ is the Higgs mass, and $\Lambda$ is the cosmological constant.
The action for the SM fermions and gauge bosons
$S_{\psi +A}\left[g_{\mu\nu}\right]$ includes the Yukawa interactions
with its dependence on the metric made explicit.
We have included the Higgs mass and the cosmological constant
since they are required as counterterms to absorb divergences, as we will see below.
We denote the beta function of a given coupling $g$ up to two-loop as
\begin{align}
	\frac{\dd g}{\dd \ln \mu} \equiv \beta_{g} = \frac{1}{\left(4\pi\right)^2}\beta_{g}^{(1)} + \frac{1}{\left(4\pi\right)^4}\beta_{g}^{(2)},
\end{align}
where $\mu$ is the renormalization scale,
and $\beta_{g}^{(1)}$ and $\beta_{g}^{(2)}$ are the one- and two-loop contributions to the beta function, respectively.

\subsection{Scalaron and conformal mode}
\label{subsec:weyl}
To see how the $R^2$-term gives rise to the scalaron degree of freedom, we first extract the conformal mode of the metric and rewrite the action in the Jordan frame~\eqref{eq:action_Jordan}. 
Without loss of generality, the metric can be decomposed as
\begin{align}
	g_{\mu\nu} = e^{2\varphi} \tilde{g}_{\mu\nu},
	\quad
	\mathrm{Det}\left[\tilde{g}_{\mu\nu}\right] = -1.
	\label{eq:metric_decomposition}
\end{align}
We call the scalar mode $\varphi$ the conformal mode of the metric;
it consists of the determinant part of the metric.
The Ricci scalar is decomposed as
\begin{align}
	R = e^{-2\varphi}\tilde{R} + 6e^{-3\varphi}\tilde{\Box}e^{\varphi},
\end{align}
where the quantities with the tilde are constructed from $\tilde{g}_{\mu\nu}$.
We redefine the fields as
\begin{align}
	H \rightarrow e^{-\varphi}H,
	\quad
	\psi \rightarrow e^{-3\varphi/2}\psi,
	\label{eq:rescale}
\end{align}
and define
\begin{align}
	\Phi \equiv \sqrt{6}M_P\,e^{\varphi}.
	\label{eq:conformal_field}
\end{align}
We also refer to $\Phi$ as the conformal mode.
Notice that the conformal mode $\Phi$ can be regarded as a \textit{measure} of dimensional quantities in units of the Planck scale.
For instance, the ratio $H/(\sqrt{6}M_P)$ is mapped to $H / \Phi$ by the rescaling of Eq.~\eqref{eq:rescale}.
Therefore, when we neglect Planck-suppressed operators in the original action, corresponding operators suppressed by $\Phi$ should be dropped after the rescaling.

By using the rescalings defined above, we may rewrite the action~\eqref{eq:action_Jordan} as
\begin{align}
	S = \int \dd^4 x &\Bigg\{ \frac{\tilde R}{12} \left( \Phi^2 + 12 \xi \abs{H} ^2\right) 
	- \frac{1}{2} \tilde{g}_{\mu\nu} \partial_\mu \Phi \partial_\nu \Phi
	+ \tilde{g}_{\mu\nu} D_\mu H^\dag D_\nu H
	- \lambda \abs{H}^4 - \frac{\lambda_m}{2} \Phi^2 \abs{H}^2 
	- \frac{\lambda_\Lambda}{4} \Phi^4 \nonumber\\
	&
	+ \left[ \left( 6 \xi + 1 \right) \abs{H}^2 + 12 \alpha \tilde R \right] \frac{\Box \Phi}{\Phi} 
	+ \alpha \tilde{R}^2 + 36 \alpha \left( \frac{\Box \Phi}{\Phi} \right)^2
	\Bigg\}
	+ S_{\psi + A} \left[ \tilde g_{\mu\nu} \right],
	\label{eq:action_jordan_2}
\end{align}
where we have defined
\begin{align}
	\lambda_m \equiv \frac{m^2}{3M_P^2},
	\quad
	\lambda_\Lambda \equiv \frac{\Lambda}{9M_P^4}.
\end{align}
An important feature is that the last term in the curly brackets involves higher derivatives, $(\Box \Phi)^2 / \Phi^2$.
Since the conformal mode $\Phi$ has a kinetic term with the wrong sign,\footnote{
	Note that the wrong-sign kinetic term of $\Phi$ is harmless because of a residual gauge symmetry~\cite{Ema:2020zvg}, which is analogous to the Coulomb potential in U$(1)$ gauge theory.
}
this higher-derivative term implies the existence of an additional \textit{physical} degree of freedom which has a kinetic term with the correct sign.

To extract this physical degree of freedom as a fundamental field, we introduce an auxiliary field $\sigma$  as
\begin{align}
	S = \int \dd^4 x &\Bigg\{ \frac{\tilde R}{12} \left( \Phi^2 + 12 \xi \abs{H} ^2\right) 
	- \frac{1}{2} \tilde{g}_{\mu\nu} \partial_\mu \Phi \partial_\nu \Phi
	+ \tilde{g}_{\mu\nu} D_\mu H^\dag D_\nu H
	- \lambda \abs{H}^4 - \frac{\lambda_m}{2} \Phi^2 \abs{H}^2 
	- \frac{\lambda_\Lambda}{4} \Phi^4 \nonumber\\
	&
	+ \left[ \left( 6 \xi + 1 \right) \abs{H}^2 + 12 \alpha \tilde R \right] \frac{\Box \Phi}{\Phi} 
	+ \alpha \tilde{R}^2 
	+ 36 \alpha \left[\left( \frac{\Box \Phi}{\Phi} \right)^2
		- \left( \frac{\Phi \sigma}{72 \alpha} + \frac{\Box \Phi}{\Phi} \right)^2
	\right]
	\Bigg\}
	+ S_{\psi + A} \left[ \tilde g_{\mu\nu} \right].
\end{align}
By substituting the solution of the constraint equation $\delta S / \delta \sigma = 0$, namely $\sigma = - 72 \alpha \Box \Phi / \Phi^2$, one recovers the original action~\eqref{eq:action_jordan_2}.
Now we can remove all the terms involving $\Box \Phi/ \Phi$ by shifting the auxiliary field $\sigma$ as
\begin{align}
	\sigma \to \sigma + \left( 6 \xi + 1 \right) \frac{\abs{H}^2}{\Phi} + 12 \alpha \frac{\tilde R}{\Phi}.
\end{align}
We then remove the kinetic mixing between $\Phi$ and $\sigma$ by the shift
\begin{align}
	\Phi \to \Phi + \sigma,
\end{align}
and we arrive at
\begin{align}
	S = \int \dd^4 x &\Bigg\{
		\frac{\tilde R}{12} \left( \Phi^2 - 2 \abs{H}^2 - \sigma^2 \right) 
		- \frac{1}{2} \tilde g^{\mu\nu} \partial^\mu \Phi \partial^\nu \Phi
		+ \tilde g^{\mu\nu} D_\mu H^\dag D_\nu H
		+ \frac{1}{2} \tilde g^{\mu\nu} \partial_\mu \sigma \partial_\nu \sigma
		 \nonumber \\
		&- \lambda \abs{H}^4 - \frac{\lambda_m}{2} \left( \Phi + \sigma \right)^2 \abs{H}^2
		- \frac{\lambda_\Lambda}{4} \left( \Phi + \sigma \right)^4
		- \frac{\lambda_\alpha}{4} \left[ \sigma \left( \Phi + \sigma \right) + 2 \bar\xi \abs{H}^2 \right]^2
	\Bigg\}
	+ S_{\psi + A} \left[ \tilde g_{\mu\nu} \right],
		\label{eq:action_conformal}
\end{align}
where we define
\begin{align}
	\lambda_{\alpha} \equiv \frac{1}{36\alpha},
	\quad
	\bar{\xi} \equiv \frac{6\xi+1}{2}.
\end{align}
This action is the starting point of our computation.
Note that the kinetic terms of the scalar fields and the scalaron $\sigma$ are canonical
and the potential contains only renormalizable terms.
This feature is related to the fact that this theory is unitary up to infinite energy if we ignore the spin-2 part,
or more generally the renormalizability of the quadratic gravity~\cite{Weinberg:1974tw,Deser:1975nv,Stelle:1976gc,Barvinsky:2017zlx,Salvio:2018crh},
as emphasized in Ref.~\cite{Ema:2020zvg}.
In the next subsection~\ref{subsec:gauge}, 
we discuss the interactions of the graviton, the gauge fixing condition and the resultant Faddeev-Popov ghost.
There we confirm that the contributions from the graviton can be safely neglected below the Planck scale
even after introducing the Faddeev-Popov ghost by properly choosing the gauge fixing condition.

Before closing this section, let us illustrate how one may write down Eq.~\eqref{eq:action_conformal} in a more familiar form.
By using the redefinitions
\begin{align}
	H \to e^{\varphi_C} H, \quad \psi \to e^{3 \varphi_C /2} \psi, \quad \sigma \to e^{\varphi_C} \sigma, \quad
	\Phi \to \sqrt{6} M_P e^{\varphi_C}, \quad
	\tilde g_{\mu\nu} \to e^{-\varphi_C}g_{\mu\nu},
\end{align}
one finds
\begin{align}
	S = \int \dd^4 x \sqrt{-g} &\Bigg\{\frac{M_P^2}{2}R \Bigg(1 - \frac{2\abs{H}^2 + \sigma^2}{6M_P^2}\Bigg) 
	+ g^{\mu\nu}D_\mu H^\dagger D_\nu H
	+ \frac{1}{2}g^{\mu\nu}\partial_\mu \sigma \partial_\nu \sigma
	- m^2\left(1+\frac{\sigma}{\sqrt{6}M_P}\right)^2\abs{H}^2 - \lambda \abs{H}^4 
	\nonumber \\ &
	- \Lambda\left(1+\frac{\sigma}{\sqrt{6}M_P}\right)^4
	- \frac{1}{144\alpha}\Bigg[\frac{3M_P^2}{2}-\Bigg(\sigma+\frac{\sqrt{6}M_P}{2} \Bigg)^2
	- \left(6\xi + 1\right)\abs{H}^2 \Bigg]^2
	\Bigg\}
	+ S_{\psi + A}\left[g_{\mu\nu}\right].
	\label{eq:action_conformal_2}
\end{align}
This is nothing but the action of the Higgs-$R^2$ theory in the conformal frame, where the Higgs and the scalaron have the conformal coupling to gravity.
Note that the rescaling factor $\varphi_C$ is different from $\varphi$ in Eq.~\eqref{eq:metric_decomposition}.

\subsection{Graviton, gauge fixing and decoupling of Faddeev-Popov ghost}
\label{subsec:gauge}

In this section, we discuss the gauge fixing condition used in this paper.
We expand the fields around the flat spacetime metric $\eta_{\mu\nu}$ as
\begin{align}
	\tilde{g}_{\mu\nu} &= [e^{h}]_{\mu\nu}
	= \eta_{\mu\rho}\left(\delta_{\nu}^{\rho} + {h^{\rho}}_{\nu} + \frac{1}{2}{h^\rho}_{\alpha} {h^\alpha}_{\nu} + \cdots\right), \label{eq:expmetric} \\
	{h^{\mu}}_{\mu} &= 0,
	\label{eq:traceless}
\end{align}
where the contractions are taken with respect to $\eta_{\mu\nu}$.\footnote{
	Remember that the beta functions do not depend on the choice of the background
	around which one expands the fields, since they are related to the UV properties of the theory. 
} The second equation follows from $\mathrm{Det}\left[\tilde{g}_{\mu\nu}\right] = -1$.

Under the general coordinate transformation $x^\mu \rightarrow x^\mu - \xi^\mu$ at the first order in $\xi^\mu$, the fields $\varphi$ and $h_{\mu\nu}$ transform as follows (see App.~\ref{app:general_coord_transf} for the derivation):\footnote{
	This $\xi^\mu$ should not be confused with the non-minimal coupling $\xi$.
}
\begin{align}
	\varphi &\rightarrow \varphi + \xi^\alpha \partial_\alpha \varphi + \frac{1}{4}\partial_\alpha \xi^\alpha, 
	\label{eq:varphi_transf} \\
	h_{\mu\nu} 
	&\rightarrow
	h_{\mu\nu} 
	+ \left[\left(\frac{\mathrm{ad}_h}{e^{\mathrm{ad}_h} -1}\right) \partial \xi\right]_{\mu\nu}
	+ \left[\left(\frac{\mathrm{ad}_h}{e^{\mathrm{ad}_h} -1}\right) \partial \xi\right]_{\nu\mu}
	- \frac{1}{2}\left(\partial_\alpha \xi^\alpha\right)\eta_{\mu\nu}
	+ \xi^\alpha \partial_\alpha h_{\mu\nu},
	\label{eq:tensor_transf}
\end{align}
where $[\partial \xi]_{\mu\nu} \equiv \partial_\mu \xi_\nu$ and
\begin{align}
	 \left[\left(\frac{\mathrm{ad}_h}{e^{\mathrm{ad}_h} -1}\right) X \right]_{\mu\nu}
	 \equiv X_{\mu\nu}
	 -\frac{1}{2}\left[h, X\right]_{\mu\nu}
	 + \frac{1}{12}\left[h, \left[h, X\right]\right]_{\mu\nu}
	 + \cdots,
	 \label{eq:adjoint}
\end{align}
with the indices raised/lowered/contracted by $\eta_{\mu\nu}$.
Here the ``ad'' is an abbreviation of the adjoint action,
and higher order terms in the right-hand-side of Eq.~\eqref{eq:adjoint} should be derived from the Taylor expansion.
Note that we have \textit{not} performed any expansion with respect to $\varphi$ nor $h_{\mu\nu}$ here.

We fix this gauge degree of freedom as
\begin{align}
	\partial^\mu h_{\mu\nu} = 0,
	\label{eq:gauge_fix}
\end{align}
in this paper.
According to the Faddeev-Popov procedure, we have to introduce ghost fields associated with this gauge fixing condition. 
Here, a crucial property of Eq.~\eqref{eq:tensor_transf} 
is that it does not depend on $\varphi$.
It follows that the Faddeev-Popov ghost associated with the gauge fixing condition~\eqref{eq:gauge_fix}
does not directly couple to the conformal mode $\varphi$.
Thus, if we can neglect contributions of the graviton $h_{\mu\nu}$,
we can also neglect contributions of the Fadeev-Popov ghost.

Now we discuss the condition under which we can neglect contributions of $h_{\mu\nu}$ 
to the RGEs. 
One may already infer from Eq.~\eqref{eq:action_conformal_2}
that $h_{\mu\nu}$ couples to the matter fields only via Planck-suppressed operators
and hence can be ignored below $M_P$,
in particular if $H$ and $\sigma$ do not develop any vacuum expectation values (VEVs),
which we assume to be the case in our computation.
On the contrary, the conformal mode of the metric and the scalaron couple to the matter fields via, \textit{e.g.}, 
$\lambda_\alpha$ and $\bar{\xi}$ that are not suppressed by $M_P$, 
and thus cannot be ignored especially for $\xi \gg 1$.
In the following, we show explicitly that this expectation is indeed the case by clarifying the meaning of physical scales in Eq.~\eqref{eq:action_conformal}.
We also allow both $H$ and $\sigma$ to have finite VEVs, as they do during inflation.

By expanding the metric $\tilde g_{\mu\nu}$ as in Eq.~\eqref{eq:expmetric} and applying the gauge fixing condition of Eq.~\eqref{eq:gauge_fix}, we obtain the kinetic term of the graviton:
\begin{align}
	\int \dd^4 x \frac{\tilde R}{12} \left( \Phi^2 - 2 \abs{H}^2 - \sigma^2 \right)
	= \int \dd^4 x \frac{1}{48} \left(  \Phi^2 - 2 \abs{H}^2 - \sigma^2 \right) \partial_\rho h_{\mu\nu} \partial^\rho h^{\mu\nu} + \mathcal{O}(h^3).
\end{align}
The graviton couples to other fields through their kinetic terms as
\begin{align}
		\tilde g^{\mu\nu} \left( - \frac{1}{2} \partial_\mu \Phi \partial_\nu \Phi + \frac{1}{2} \partial_\mu \sigma \partial_\nu \sigma + D_\mu H^\dag D_\nu H \right)
		\supset - h^{\mu\nu} \left( - \frac{1}{2} \partial_\mu \Phi \partial_\nu \Phi +  \frac{1}{2} \partial_\mu \sigma \partial_\nu \sigma +  D_\mu H^\dag D_\nu H \right) + \mathcal{O} (h^2).
\end{align}
After canonically normalizing the graviton, one may see that the coupling of the graviton to the matter fields and the conformal mode is suppressed by $\sqrt{\Phi^2 - 2 \abs{H}^2 - \sigma^2}$.
Hence, at a given energy scale $\mu$, we can safely neglect the coupling to $h_{\mu\nu}$ 
as long as 
\begin{align}
	\frac{\mu}{\sqrt{\Phi^2 - 2 \abs{H}^2 - \sigma^2}} \ll 1,
	\label{eq:condition_graviton_neglected}
\end{align}
which we require throughout this paper.
The physical meaning of this condition is as follows.
If the scalaron $\sigma$ and the Higgs field $H$ do not develop VEVs comparable to $\Phi$, one can  simplify this condition as $\mu / \Phi \ll 1$. This implies
that a typical scale in the conformal frame \eqref{eq:action_conformal_2}, $\mu_\text{C} \equiv e^{-\varphi_C} \mu$,\footnote{
	For a given metric $g_{\mu\nu}$, its typical scale $\mu$ is defined as $\mu = (g_{\mu\nu} \Delta x^\mu \Delta x^\nu)^{-1/2}$ with $\Delta x^\mu$ being a typical length scale.
	Let the typical scale in $\tilde g_{\mu\nu}$ be $\mu$.
	Then, the typical scale $\mu_\bullet$ in a metric of $g_{\mu\nu} = e^{\varphi_\bullet} \tilde g_{\mu\nu}$ can be expressed as $\mu_\bullet = e^{- \varphi_\bullet} \mu$.
} should be below the Planck scale $\mu_\text{C} / M_P \ll 1$.
However, during inflation, $\sigma$ and $H$ develop their VEVs, and in particular the VEV of $\sigma$ becomes comparable to $\Phi$, which enhances the coupling to the graviton.
To understand the physical meaning of the condition in this case, let us recall how the action in Eq.~\eqref{eq:action_conformal} is related to that in the Einstein frame.
The action in the Einstein frame is obtained by the following redefinitions
\begin{align}
	\Phi_E^2 = \Phi^2 - 2 \abs{H}^2 - \sigma^2, \quad \Phi_E = \sqrt{6} M_P e^{\varphi_E}, \quad
	g_{\mu\nu} =  e^{2 \varphi_E} \tilde g_{\mu\nu},
\end{align}
and an appropriate rescaling of the fields (see App.~\ref{app:eins-conf} for details).
Now the condition~\eqref{eq:condition_graviton_neglected} 
can be written as 
\begin{align}
	\frac{\mu_\text{E}}{M_P} \ll 1,
\end{align}
with $\mu_E \equiv e^{\varphi_E} \mu$ being a typical scale in the Einstein frame.

From these observations, we can simplify our computation of the RGEs as follows.
\begin{enumerate}

\item
We can ignore $h_{\mu\nu}$ as long as the condition~\eqref{eq:condition_graviton_neglected} is met
for a given scale $\mu$.
This condition implies that a typical energy scale in the Einstein frame should be below the Planck scale, \emph{i.e.}, $\mu_E / M_P \ll 1$.

\item
We do not have to take into account any Faddeev-Popov ghosts,
with the assumption that our gauge fixing condition is Eq.~\eqref{eq:gauge_fix},
once we can ignore $h_{\mu\nu}$.

\end{enumerate}
Thus, we take the metric as $\tilde{g}_{\mu\nu} = \eta_{\mu\nu}$ to compute the RGEs below $M_P$ in the following.
The action then reduces to
\begin{align}
	S
	= \int \dd^4 x &\left[
	- \frac{1}{2}\eta^{\mu\nu} \partial_\mu \Phi \partial_\nu \Phi
	+ \eta^{\mu\nu}D_\mu H^\dagger D_\nu H
	+ \frac{1}{2}\eta^{\mu\nu}\partial_\mu \sigma \partial_\nu \sigma
	- \lambda \abs{H}^4 
	\right. \nonumber \\ &\left.
	- \frac{\lambda_m}{2}\left(\Phi + \sigma \right)^2\abs{H}^2
	- \frac{\lambda_\Lambda}{4}\left(\Phi+\sigma\right)^4
	- \frac{\lambda_\alpha}{4}\left(\sigma \left(\Phi + \sigma\right)
	+ 2\bar{\xi} \abs{H}^2 \right)^2
	\right]
	+ S_{\psi + A}\left[\eta_{\mu\nu}\right].
	\label{eq:LSM}
\end{align}
It can be viewed as a LSM composed of $\Phi$, $\sigma$ and $H$
coupled to the SM fermions and gauge bosons,
as studied in detail in Ref.~\cite{Ema:2020zvg}.\footnote{
 	The inclusion of $\Phi$ makes this definition of the LSM invariant under a frame transformation,
	\textit{i.e.}, the Weyl transformation.
}
This LSM contains only two additional scalar modes, the conformal mode $\Phi$ and the scalaron,
on top of the SM particles,
which greatly simplifies our computation of the RGEs.
An important feature is that the kinetic term of the conformal mode $\Phi$ 
has a wrong sign and hence this field is ghost-like.
It is still harmless thanks to a residual gauge symmetry
(see Ref.~\cite{Ema:2020zvg} for more details),
playing a role that is similar to the Coulomb potential of U$(1)$ gauge theory.
In the following,
we derive the one- and two-loop RGEs of the Higgs-$R^2$ model 
from this action in Secs.~\ref{subsec:one_loop} and~\ref{subsec:two_loop}, respectively.
Of the SM parameters we include only the top Yukawa $y_t$ and the gauge couplings $g_3, g_2$ and $g_1$ in the RGEs, 
since the other couplings are all much smaller than unity.
See Apps.~\ref{app:one_loop} and~\ref{app:two_loop} for details of the derivation.

\subsection{One-loop RGE}
\label{subsec:one_loop}
In this subsection, we give the one-loop RGEs of the Higgs-$R^2$ theory
that are valid up to $M_P$.
At the one-loop level, the RGEs are given by 
\begin{align}
	\beta_{g_1}^{(1)} &= \frac{41}{10}g_1^3,
	\quad
	\beta_{g_2}^{(1)} = -\frac{19}{6}g_2^3,
	\quad
	\beta_{g_3}^{(1)} = -7g_3^3, \\
	\beta_{y_t}^{(1)} &= y_t\left[\frac{9 y_t^2}{2}-\frac{17}{20} g_1^2-\frac{9}{4} g_2^2-8 g_3^2\right], \\
	\beta_{\lambda}^{(1)} 
	&= \left(8\bar{\xi}^2 - 8 \bar{\xi} + 2\right)\bar{\xi}^2 \lambda_\alpha^2
	+24 \bar{\xi }^2 \lambda  \lambda _{\alpha } 
	+24\lambda ^2-6 y_t^4
	+\frac{27 g_1^4}{200}
	+\frac{9 g_2^4}{8}+\frac{9}{20} g_1^2 g_2^2
	+ \left[12  y_t^2-\frac{9 g_1^2}{5}-9 g_2^2\right]\lambda,
	\label{eq:beta_lambda} \\
	\beta_{\lambda_m}^{(1)}
	&= 2\bar{\xi}\left(2 \bar{\xi }-1\right) \lambda _{\alpha }^2
	-8 \bar{\xi }\lambda_m^2
	+ \lambda_m\left[4 \bar{\xi }^2 \lambda _{\alpha }+8 \bar{\xi } \lambda _{\alpha }-3 \lambda _{\alpha }
	+12 \lambda+6 y_t^2
	-\frac{9 g_1^2}{10}-\frac{9 g_2^2}{2}\right], \\
	\beta_{\bar{\xi}}^{(1)}
	&= \bar{\xi}\left[\left(4\bar{\xi}^2 + 4\bar{\xi} - 3\right)  \lambda_\alpha
	 +12 \lambda +6  y_t^2
	-\frac{9}{10} g_1^2 -\frac{9}{2} g_2^2\right], \\
	\beta_{\lambda_\alpha}^{(1)}
	&=\left(8 \bar{\xi }^2+5\right) \lambda _{\alpha }^2, \\
	\beta_{\lambda_\Lambda}^{(1)}
	&=
	\frac{\lambda _{\alpha }^2}{2} - 2 \lambda _{\alpha } \lambda _{\Lambda }
	-16 \bar{\xi } \lambda _{\Lambda } \lambda _m +2 \lambda _m^2.
\end{align}
In App.~\ref{app:one_loop}, we have derived these RGEs by computing scalar four-point functions.
In the course of the derivation, 
we have explicitly checked that all the divergences that appear at the one-loop level 
are indeed renormalized by the operators within the action~\eqref{eq:LSM}.
It is a non-trivial verification of our formulation based on the decomposition~\eqref{eq:metric_decomposition}
and the gauge fixing condition~\eqref{eq:gauge_fix}.
We have also checked that these RGEs agree with the results obtained with the help of 
\texttt{SARAH}~\cite{Staub:2013tta}
outlined in App.~\ref{app:two_loop}.
Moreover, these RGEs coincide with Refs.~\cite{Salvio:2014soa,Salvio:2017qkx,Salvio:2018crh}
if we ignore the contributions from the spin-2 particles in the latter.\footnote{
	Some pieces of the RGEs shown here have also been derived, \textit{e.g.}, in Refs.\cite{Fradkin:1981iu,Buchbinder:1992rb,Elizalde:1993ee,Elizalde:1993ew,Codello:2015mba,Markkanen:2018bfx,Kubo:2018kho}.
	However, Refs.~\cite{Buchbinder:1992rb,Elizalde:1993ee,Elizalde:1993ew,Codello:2015mba} obtained the opposite sign for $\beta_\alpha$.
}

Several comments are in order. 
First, it can be seen that the Higgs mass and the cosmological constant emerge due to the RG running 
even if they are highly suppressed at a lower energy scale.
This is simply an example of the infamous fine tuning problem;
the scalaron introduces an additional scale, the scalaron mass scale.
The Higgs mass and the cosmological constant are sensitive to this additional scale
since they are not protected by any symmetry.
Indeed, in order to realize the current universe with the electroweak (EW) 
scale and the almost vanishing cosmological constant,
we have to tune the UV boundary conditions such that
\begin{align}
	\lambda_m(\mu = m_\sigma) \sim \frac{v_\mathrm{EW}^2}{m_\sigma^2}\lambda_\alpha
	\ll \lambda_\alpha,
	\label{eq:bc_m_Lambda}
\end{align}
and a similar condition for the cosmological constant,\footnote{
	The cosmological constant may be affected by the EW scale
	below the scalaron mass scale,
	and hence the condition to realize the current tiny cosmological constant
	may need to take this into account.
}
where $m_\sigma^2 = M_P^2/12\alpha$ is the scalaron mass scale and
$v_\mathrm{EW} = 246\,\mathrm{GeV}$.
It is indeed a tuning since the natural size of $\lambda_m$
inferred from the RGEs is $\lambda_m \sim \lambda_\alpha$
for $\xi \gg 1$.
These massive parameters were not included in the previous studies on
the inflationary prediction of this model. 
Thus, we study their effects on the inflationary prediction
in Sec.~\ref{sec:inf}.

Second, our method allows us to compute the beta functions of only the ratios of the massive parameters,
not the massive parameters themselves.
This is, however, not a limitation of our method, since only the ratios are important for physics.\footnote{
	One can see this fact, \textit{e.g.}, by remembering that one can compute the spectral index and the tensor-to-scalar ratio
	without any problem even in Planck units $M_P = 1$.
}
This point is also emphasized in Refs.~\cite{Salvio:2014soa,Salvio:2017qkx}.

Finally, it may be interesting to observe that a parameter with a higher mass dimension does not contribute 
to the beta functions of those with a lower mass dimension. For instance, $\lambda_\Lambda$ does not contribute
to $\beta_{\lambda_\alpha}$, $\beta_{\bar{\xi}}$, $\beta_{\lambda}$ and $\beta_{\lambda_m}$,
while $\lambda_m$ does not contribute to $\beta_{\lambda_\alpha}$, $\beta_{\bar{\xi}}$ and $\beta_{\lambda}$.
This is trivial in the original Jordan frame language, 
but it is non-trivial once we map the theory to the LSM with the renormalizable potential~\eqref{eq:LSM}.
From this point of view,
all the couplings $\lambda_\alpha$, $\lambda_\Lambda$, $\bar{\xi}$, $\lambda_m$ and $\lambda$
are equally dimensionless,
and hence it is not entirely obvious (at least to the present authors) that the above hierarchy indeed holds.

\subsection{Two-loop RGE}
\label{subsec:two_loop}
In this subsection, we show the two-loop RGEs of the Higgs-$R^2$ theory valid below the Planck scale 
obtained with the help of \texttt{SARAH}~\cite{Staub:2013tta}.
It is non-trivial to translate the outputs of \texttt{SARAH} (or public codes in general)
to the RGEs of the Higgs-$R^2$ theory since we have to take into account the ghost-like property of $\Phi$. 
We only show the final results in this subsection,
and explain how this translation can be done in App.~\ref{app:two_loop}.
We note that, during the course of deriving the results, we have performed several non-trivial checks,
which strongly supports the validity of our formalism. We again refer to App.~\ref{app:two_loop} on this point.

The RGEs of the gauge couplings are not affected by $\Phi$ nor $\sigma$ up to two-loop,
and are given by
\begin{align}
	\beta_{g_1}^{(2)} &= 
	g_1^3 \left[ \frac{199}{50}g_1^2 + \frac{27}{10} g_2^2 + \frac{44}{5}g_3^2 - \frac{17}{10}y_t^2\right], \\
	\beta_{g_2}^{(2)} &= 
	g_2^3 \left[ \frac{9}{10}g_1^2 + \frac{35}{6} g_2^2 + 12g_3^2 - \frac{3}{2}y_t^2\right], \\
	\beta_{g_3}^{(2)} &= 
	g_3^3 \left[ \frac{11}{10}g_1^2 + \frac{9}{2} g_2^2 -26g_3^2 - 2y_t^2\right].
\end{align}
The RGE of the top Yukawa coupling is affected by $\Phi$ and $\sigma$ and becomes
\begin{align}
	\beta_{y_t}^{(2)}
	&= -12 y_t^5 + y_t^3\left[-12 \bar{\xi }^2 \lambda _{\alpha }+\frac{393 g_1^2}{80}+\frac{225 g_2^2}{16}+36 g_3^2-12 \lambda\right]
	\nonumber \\
	&+ y_t \left[6 \bar{\xi }^4 \lambda _{\alpha }^2+\frac{1}{2} \bar{\xi }^2 \lambda _{\alpha }^2
	+12 \bar{\xi }^2 \lambda\lambda _{\alpha}
	+\frac{1187 g_1^4}{600}-\frac{23 g_2^4}{4}-108 g_3^4-\frac{9}{20} g_1^2 g_2^2+\frac{19}{15} g_1^2 g_3^2+9 g_2^2 g_3^2+6 \lambda ^2\right].
\end{align}
The RGE of the Higgs quartic coupling is given by

\begin{align}
	\beta_{\lambda}^{(2)}
	&= -312\lambda^3 
	+ \lambda^2\left[-816 \bar{\xi }^2 \lambda _{\alpha }+\frac{108 g_1^2}{5}+108 g_2^2-144 y_t^2\right]
	+\lambda y_t^2\left[-144 \bar{\xi }^2 \lambda _{\alpha }+\frac{17 g_1^2}{2}+\frac{45 g_2^2}{2}+80 g_3^2\right]
	\nonumber \\
	&+ \lambda \left[-696 \bar{\xi }^4 \lambda _{\alpha }^2+144 \bar{\xi }^3 \lambda _{\alpha }^2
	-10 \bar{\xi }^2 \lambda _{\alpha }^2+\frac{72}{5} g_1^2\bar{\xi }^2 \lambda _{\alpha }
	+72 g_2^2 \bar{\xi }^2 \lambda _{\alpha }+\frac{1887 g_1^4}{200}-\frac{73 g_2^4}{8}+\frac{117}{20}g_1^2 g_2^2\right]
   	\nonumber \\
	&+ y_t^4 \left[24 \bar{\xi }^2 \lambda _{\alpha }-\frac{8 g_1^2}{5}-32 g_3^2\right]
	+ y_t^2 \left[-48 \bar{\xi }^4 \lambda _{\alpha }^2+24 \bar{\xi }^3 \lambda _{\alpha }^2
	-\frac{171 g_1^4}{100}-\frac{9 g_2^4}{4}+\frac{63}{10} g_1^2g_2^2\right]
	\nonumber \\
	&-192 \bar{\xi }^6 \lambda _{\alpha }^3+112 \bar{\xi }^5 \lambda _{\alpha }^3-8 \bar{\xi }^4 \lambda _{\alpha }^3+20 \bar{\xi }^3
   \lambda _{\alpha }^3-10 \bar{\xi }^2 \lambda _{\alpha }^3+\frac{12}{5} g_1^2 \bar{\xi }^4 \lambda _{\alpha }^2+12 g_2^2 \bar{\xi }^4
   \lambda _{\alpha }^2-\frac{6}{5} g_1^2 \bar{\xi }^3 \lambda _{\alpha }^2-6 g_2^2 \bar{\xi }^3 \lambda _{\alpha }^2
   	\nonumber \\
   	&+\frac{27}{25}g_1^4 \bar{\xi }^2 \lambda _{\alpha }+9 g_2^4 \bar{\xi }^2 \lambda _{\alpha }
	+\frac{18}{5} g_1^2 g_2^2 \bar{\xi }^2 \lambda _{\alpha}-\frac{3411 g_1^6}{2000}+\frac{305 g_2^6}{16}-\frac{289}{80} 
	g_1^2 g_2^4-\frac{1677}{400} g_1^4 g_2^2.
\end{align}
The RGEs of the Higgs mass term and the non-minimal coupling are given by
\begin{align}
	\beta_{\lambda_m}^{(2)}
	&= \lambda_m^2 \left[48 \bar{\xi }^2 \lambda _{\alpha }+20 \bar{\xi } \lambda _{\alpha }
	-\frac{48}{5} g_1^2 \bar{\xi }-48 g_2^2 \bar{\xi }+48 \bar{\xi }y_t^2\right]
	+ \lambda_m \lambda_\alpha^2 \left[-60 \bar{\xi }^4-80 \bar{\xi }^3+9 \bar{\xi }^2-16 \bar{\xi }+23\right]
	\nonumber \\
	&+ \lambda_m \lambda_\alpha\left[\frac{24}{5} g_1^2 \bar{\xi }^2+24 g_2^2 \bar{\xi }^2
	+\frac{6}{5} g_1^2 \bar{\xi }+6 g_2^2 \bar{\xi }-120 \lambda  \bar{\xi }^2
	-144\lambda  \bar{\xi }-24 \bar{\xi }^2 y_t^2-24 \bar{\xi } y_t^2\right]
	\nonumber \\
	&+ \lambda_m\left[\frac{72 g_1^2 \lambda }{5}+72 g_2^2 \lambda +\frac{17}{4} g_1^2 y_t^2+\frac{45}{4} g_2^2 y_t^2
	+40 g_3^2 y_t^2
	+\frac{1671g_1^4}{400}+\frac{9}{8} g_2^2 g_1^2
	-\frac{145 g_2^4}{16}-60 \lambda ^2-72 \lambda  y_t^2-\frac{27 y_t^4}{2}\right]
	\nonumber \\
	&+\lambda_\alpha^3 \left[-56 \bar{\xi }^4-22 \bar{\xi }^2+18 \bar{\xi }\right]
	+ \lambda_\alpha^2 \left[\frac{3}{5} g_1^2 \bar{\xi }^2+3 g_2^2 \bar{\xi }^2-72 \lambda  \bar{\xi }^2-12 \bar{\xi }^2 y_t^2\right],
\end{align}
and
\begin{align}
	\beta_{\bar{\xi}}^{(2)}
	&= -60 \lambda _{\alpha }^2\bar{\xi}^5 -40 \lambda _{\alpha }^2 \bar{\xi}^4
	+ \bar{\xi}^3\left[9 \lambda _{\alpha }^2-120 \lambda  \lambda _{\alpha }+\frac{24}{5} g_1^2 \lambda _{\alpha }
	+24 g_2^2 \lambda _{\alpha }-24 \lambda_{\alpha } y_t^2\right]
	\nonumber \\
	&+ \bar{\xi}^2\left[-12 \lambda _{\alpha }^2-72 \lambda  \lambda _{\alpha }+\frac{3}{5} g_1^2 \lambda _{\alpha }
	+3 g_2^2 \lambda _{\alpha }-12 \lambda_{\alpha } y_t^2\right]
	\nonumber \\
	&+ \bar{\xi}\left[\frac{25 \lambda _{\alpha }^2}{2}+\frac{72 g_1^2 \lambda }{5}+72 g_2^2 \lambda +\frac{17}{4} g_1^2 y_t^2
	+\frac{45}{4} g_2^2 y_t^2+40g_3^2 y_t^2
	+\frac{1671 g_1^4}{400}+\frac{9}{8} g_2^2 g_1^2-\frac{145 g_2^4}{16}-60 \lambda ^2-72 \lambda  y_t^2-\frac{27 y_t^4}{2}\right].
\end{align}
Finally, the RGEs of $\lambda_\alpha$ and $\lambda_\Lambda$ are given by
\begin{align}
	\beta_{\lambda_\alpha}^{(2)}
	&= 
	\frac{1}{5} \lambda _{\alpha }^2 \left[48 \bar{\xi }^2 \left(g_1^2+5 g_2^2-5 y_t^2\right)-5 \left(32 \bar{\xi }^3+20 \bar{\xi
   }^2+15\right) \lambda _{\alpha }\right],
\end{align}
and

\begin{align}
	\beta_{\lambda_\Lambda}^{(2)}
	&=
	\lambda_\alpha^3\left[-4 \bar{\xi }^2-7\right]
	+ \lambda_\alpha^2 \left[32 \bar{\xi }^3 \lambda _{\Lambda }+20 \bar{\xi }^2 \lambda _{\Lambda }
	-8 \bar{\xi }^2 \lambda _m-8 \bar{\xi } \lambda _m+26 \lambda_{\Lambda }\right]
	\nonumber \\
	&+ \lambda_\alpha\left[96 \bar{\xi }^2 \lambda _{\Lambda } \lambda _m+40 \bar{\xi } \lambda _{\Lambda } \lambda _m
	-16 \bar{\xi } \lambda _m^2-4 \lambda _m^2\right]
	\nonumber \\
	&-\frac{96}{5} g_1^2 \bar{\xi } \lambda _{\Lambda } \lambda _m-96 g_2^2 \bar{\xi } \lambda _{\Lambda } \lambda _m
	+96 \bar{\xi } \lambda_{\Lambda } \lambda _m y_t^2+\frac{12}{5} g_1^2 \lambda _m^2+12 g_2^2 \lambda _m^2
	-12 \lambda _m^2 y_t^2.
\end{align}
These are the full two-loop RGEs of the Higgs-$R^2$ theory
that are valid below the Planck scale.
To our knowledge, it is this paper that has derived them for the first time.

\section{Inflation with massive parameters}
\label{sec:inf}
In Sec.~\ref{sec:rge}, we have seen that the Higgs mass and the cosmological constant
arise due to the RG running above the scalaron mass scale 
even if they are tiny at low energies.
In this section, we study whether these massive parameters change the inflationary prediction of 
the Higgs-$R^2$ model or not.
We assume here that the Higgs quartic coupling $\lambda$ is positive
due to either an uncertainty in the value of the top quark mass or some new physics.
In Sec.~\ref{sec:quartic}, we study effects of the scalaron on the sign of $\lambda$ at high energy in detail.
We assume that the values of $\alpha$, $m$ and $\Lambda$ are consistent with 
the values naturally inferred from the RGE's, \emph{i.e.}
\begin{align}
	\alpha \sim \xi^2\, , \quad \frac{m}{M_P} \sim \frac{1}{\xi}\, ,\quad \frac{\Lambda}{M_P^4} \sim \frac{1}{\xi^4}.\label{eq:estpars}
\end{align}
We also assume that $\xi \gg 1$ in our analysis,
as in the typical case of Higgs-$R^2$ inflation.

\subsection{Background evolution}
The inflationary trajectory is most easily studied in the Einstein frame. Starting with the action of Eq.~\eqref{eq:action_Jordan}, 
we introduce an auxiliary field $\chi$ to rewrite it as
\begin{align}
	S =& \int \dd^4 x \sqrt{-g} 
	\left[\frac{M_P^2}{2}R\left(1+\frac{2\xi \abs{H}^2 + 4\chi}{M_P^2}\right) 
	+ g^{\mu\nu}D_\mu H^\dagger D_\nu H
	- m^2\abs{H}^2 - \lambda \abs{H}^4 - \frac{ \chi^2}{\alpha} - \Lambda
	\right],
\end{align}
where we neglected the interactions with the SM fermions and gauge fields, 
as they do not affect the inflationary trajectory.
We make the following Weyl transformation
\begin{align}
	g_{\mu\nu} \rightarrow \Omega_E^{-2} g_{\mu\nu}\,, \quad \Omega_E^2 = 1 + \frac{2\xi |H|^2 + 4\chi }{M_P^2}.	
\end{align}
Note that we only rescale the metric, but the other fields are unaffected. 
Upon defining the field
\begin{align}
	\frac{\sigma_E}{M_P} = \sqrt{\frac{3}{2}} \log{\Omega_E^2},	
\end{align}
the action in the Einstein frame is given by
\begin{align}
	S_E = \int \dd^4 x \sqrt{- g} \Bigg[ &\frac{M_P^2}{2} R 
	+\frac 1 2  g^{\mu\nu}\partial_\mu \sigma_E \partial_\nu \sigma_E + \frac1 2  e^{-\sqrt{\frac 2 3}\frac{\sigma_E}{M_P}}g^{\mu\nu} \partial_\mu \phi \partial_\nu \phi \nonumber
	\\& -e^{-2\sqrt{\frac 2 3}\frac{\sigma_E}{M_P}} \left(\frac{m^2}{2}\phi^2 + \frac{\lambda}{4} \phi^4 + \frac{M_P^4}{16\alpha} \left(e^{\sqrt{\frac 2 3}\frac{\sigma_E}{M_P}} - 1 - \frac{\xi \phi^2}{M_P^2} \right)^2 + \Lambda \right)
	\Bigg].
	\label{eq:action_einstein}
\end{align}
We are using unitary gauge, in which the radial direction of the Higgs field $\phi$ is defined as
\begin{align}
	H = \frac{1}{\sqrt 2}\begin{pmatrix}0 \\ \phi \end{pmatrix}.
\end{align}
The inflationary trajectory of this theory in the limit $m,\Lambda = 0$ was studied in Refs.~\cite{Ema:2017rqn, Wang:2017fuy, He:2018gyf, Gundhi:2018wyz, Enckell:2018uic}. We will now study the effect of the mass term and cosmological constant on the background evolution. Our analysis is similar to the one in Ref.~\cite{Gundhi:2018wyz}.
In the following we omit the subscript $E$ from $\sigma_E$ for notational simplicity,
but it should be noticed that it is not exactly the $\sigma$-meson of the LSM~\eqref{eq:LSM}
contrary to $\sigma$ in the previous section.

We will use $\Phi^I$ as shorthand notation for the fields
\begin{align}
	\Phi^I = \begin{pmatrix} \sigma \\  \phi \end{pmatrix}.	
\end{align}
It should not be confused with the conformal mode $\Phi$ since the former always carries an index $I$.
In our analysis in the Einstein frame, the $\phi$-field has a non-canonical kinetic term, requiring a covariant treatment of the field space \cite{Sasaki:1995aw, Peterson:2010np, Gong:2011uw, Kaiser:2012ak}.
Here let us emphasize that there are two different definitions of the ``target space" 
(or ``field space") in the literature.
In Sec.~\ref{sec:rge}, we have identified the Higgs-$R^2$ theory as a LSM.
In this identification, we have defined the target space
by including not only the scalar fields but also the conformal mode of the metric $\Phi$, 
following Ref.~\cite{Ema:2020zvg}.
The inclusion of $\Phi$ makes this definition of the target space frame independent,
and hence this definition is useful to discuss the unitarity structure and the cut-off scale 
of the theory, which are of course frame independent.
On the other hand,
the target space is often defined just by the kinetic terms of the scalar fields,
without including $\Phi$, in the context of inflationary perturbations 
as in Refs.~\cite{Sasaki:1995aw, Peterson:2010np, Gong:2011uw, Kaiser:2012ak}.
This definition of the target space is frame dependent, nevertheless it is useful to
compute inflationary perturbations.
The frame dependence does not cause any problem
as long as one works only in the Einstein frame and never moves to other frames.
In order to distinguish between the two definitions of the target space, and in order to emphasize that the second target space
is understood to be defined only in the Einstein frame, we refer to the latter as the ``Einstein frame target space" in the following.

The Einstein frame target space metric is given by
\begin{align}
	G_{IJ} = \begin{pmatrix}
	1 & 0 \\
	0 & e^{-\sqrt{\frac 2 3}\frac{\sigma}{M_P}}
	\end{pmatrix}\, .\label{eq:metricfieldspace}	
\end{align}
It induces a covariant derivative $\mathcal D_I$, which operates on a vector in the field space $V^J$ as
\begin{align}
	\mathcal D_I V^J = \frac{\partial V^J}{\partial \Phi^I} + \Gamma^J\,_{I K}V^K\, ,	
\end{align}
with $\Gamma^J\,_{IK}$ the Christoffel symbols of the Einstein frame target space. The background equation of motion of the fields $\Phi^I$ becomes
\begin{align}
	\mathcal D_t \dot \Phi^I = - 3 \mathcal{H} \dot \Phi^I - G^{IJ} V_{,J},
\end{align}
where $V$ is the potential and $\mathcal D_t = \dot\Phi^I \mathcal D_I$.
The Hubble parameter $\mathcal H$ and its evolution are given by
\begin{align}
	\mathcal{H}^2 = \frac{1}{3 M_P^2} \left(\frac 12 G_{IJ} \dot \Phi^I \dot \Phi^J + V(\Phi) \right) \, , \quad 
	\dot{ \mathcal{H}} = \frac{1}{2 M_P^2} G_{IJ} \dot \Phi^I \dot \Phi^J.
\end{align}
For our Einstein frame target space of Eq.~\eqref{eq:metricfieldspace} the nonzero Christoffel symbols are
\begin{align}
	\Gamma^{\sigma}\,_{\phi \phi} = \frac{e^{-\sqrt{\frac 2 3}\frac{\sigma}{M_P}}}{\sqrt 6 M_P}\, , \quad \Gamma^\phi\,_{\sigma  \phi} = \Gamma^\phi\,_{\phi \sigma} = -\frac{1}{\sqrt 6 M_P}, 	
\end{align}
and the background equations of motion are
\begin{align}
		& \ddot \sigma = -3 \mathcal H \dot \sigma - \frac{e^{-\sqrt \frac 2 3 \frac{\sigma}{M_P}  } }{\sqrt 6 M_P}\dot\phi^2 - V_{,\sigma} \, \label{eq:eompsi} ,\\
		& \ddot \phi = -3 \mathcal H \dot \phi + \sqrt{\frac{2}{3}}\frac{1}{M_P} \dot \phi \dot \sigma - e^{\sqrt\frac 23 \frac{\sigma}{M_P}} V_{,\phi} \, .\label{eq:eomphi}
\end{align}
We will study the inflationary prediction of this model in the following.

\subsection{Inflationary prediction}
The potential $V$ has two valleys that are symmetrically aligned with respect to the $\phi$-axis. For our parameter choice we expect inflation to take place along this valley \cite{Ema:2017rqn, Gundhi:2018wyz}. Setting $V_{,\phi}=0$ gives the $\sigma$-dependence of the $\phi$-field in the valley
\begin{align}
	\phi_v^2(\sigma) = \frac{ \xi M_P^2 \left(e^{\sqrt \frac 2 3 \frac{\sigma}{M_P}} -1 \right) -4 \alpha m^2}{\xi^2  + 4 \alpha \lambda}.\label{eq:valley}
\end{align}
We assume that during inflation $e^{\sqrt \frac 2 3 \frac{\sigma}{M_P}}\gg 1$ (and we will see that this indeed gives us a consistent solution with at least 60 e-folds of inflation). During inflation, the contribution from the mass term to $\phi_v(\sigma)$ is thus negligible for $\xi \gg 1$.

Inserting Eq.~\eqref{eq:valley} into the kinetic term for $\phi$ we can express it in terms of $\sigma$
\begin{align}
	\frac 1 2 e^{-\sqrt\frac 2 3 \frac{\sigma}{M_P}} g^{\mu\nu} \partial_\mu \phi \partial_\nu \phi \sim \frac{\xi}{12(4 \alpha\lambda + \xi^2)} g^{\mu\nu} \partial_\mu \sigma \partial_\nu \sigma,
\end{align}
where we used $e^{\sqrt \frac 2 3 \frac{\sigma_E}{M_P}}\gg1$ to obtain the RHS. The kinetic term of $\phi$ is thus suppressed by $\mathcal O(1/\xi)$ with respect to the kinetic term of $\sigma$ 
and we will neglect it in our analytical approximation. Inserting \eqref{eq:valley} into the potential gives:
\begin{align}
	V(\sigma) = \frac{e^{-2 \sqrt \frac 2 3 \frac{\sigma}{M_P}} }{4(4 \alpha \lambda + \xi^2)}
	\left[-4 m^4 \alpha + M_P^4 \lambda\left(e^{\sqrt \frac 2 3 \frac{\sigma}{M_P}} -1 \right)^2 
	  + 2m^2 M_P^2 \xi \left(e^{\sqrt \frac 2 3 \frac{\sigma}{M_P}} -1 \right)  
	  + 4 \Lambda\left(4 \alpha \lambda + \xi^2\right) \right].
\end{align}
From this potential we can compute the inflationary observables in the slow-roll approximation. Note that in the regime $e^{\sqrt \frac 2 3 \frac{\sigma}{M_P}}\gg1$ and for the parameter values of Eq.~\eqref{eq:estpars} the expression between square brackets is dominated by the term $M_P^4 \lambda \, e^{2\sqrt \frac 2 3 \frac{\sigma}{M_P}}$. 

We first determine the value of $\sigma$ at the moment when the observable modes left the horizon, $\sigma_*$, using:
\begin{align}
	N_* - N_\text{end} \sim \frac{1}{M_P^2} \int_{\sigma_\text{end}}^{\sigma_*} \dd \sigma \frac{V}{V_{,\sigma}} \sim\frac{1}{M_P^2} \int_{\sigma_\text{end}}^{\sigma_*} \dd \sigma \frac{\sqrt\frac 3 2 M_P^3 \lambda e^{\sqrt \frac 2 3 \frac{\sigma}{M_P}}}{2 M_P^2 \lambda - 2m^2 \xi}, \label{eq:noef}	
\end{align}
and we set $N_\text{end} =0$. We neglect the contribution from the lower bound of the integral in \eqref{eq:noef} and find
\begin{align}
	N_* = \frac{\frac 3 4 M_P^2 \lambda e^{\sqrt \frac 2 3 \frac{\sigma_*}{M_P}}}{M_P^2 \lambda - m^2 \xi} \quad \rightarrow \quad \sigma_* = \frac 3 2 M_P \log{\left[ \frac 4 3 \left(1 - \frac{m^2 \xi}{M_P^2 \lambda} \right) N_* \right]},
\end{align}
where we have assumed $\xi m^2 < \lambda M_P^2$,
which is the case for the typical value of $m$ as long as $\lambda \xi \gg 1$.
Plugging this value into the slow-roll parameters $\epsilon_V$ and $\eta_V$ gives:
\begin{align}
		&\epsilon_V = \frac{M_P^2}{2} \left( \frac{V_{,\sigma}}{V} \right)^2 \simeq 
		\frac{3}{4 N_*^2}\label{eq:sreps}, \\
		&\eta_V = M_P^2\frac{V_{,\sigma\sigma}}{V} \simeq - \frac{1}{N_*}\label{eq:sreta},
\end{align}
in the limit of large $N_*$. The scalar spectral index $n_\mathcal R^*$ and the tensor to scalar ratio $r^*$ have the usual values of the $R^2$-like inflationary models
\begin{align}
	n_\mathcal R^* = 1 - \frac{2}{N_*}, \quad r^* = \frac{12}{ N_*^2}.	
\end{align}
The scalar amplitude is
\begin{align}
	A_\mathcal{R}^* = \frac{V^*}{24 \pi^2 M_P^4 \epsilon^*_V} = \frac{\lambda N_*^2}{72 \pi^2 (4 \alpha\lambda + \xi^2)}.\label{eq:PSan}	
\end{align}
Choosing $N_* \sim 50 -60$, measurement of the power spectrum fixes the relation
\begin{align}
	\frac{\xi^2}{\lambda} + 4 \alpha \sim 2 \times 10^9.	
\end{align}
Our analytical analysis above shows that none of the inflationary observables is affected by the presence of the nonzero Higgs mass term and cosmological constant.
In other words, these massive parameters, inevitably induced due to the RGEs, 
do not spoil the consistency of Higgs-$R^2$ inflation with the CMB observation.

As a consistency check we will compute the turn rate in the valley, 
which ought to remain small for the single-field approximation to be valid. 
We introduce unit vectors $\hat T$ and $\hat N$ 
that are covariant with respect to coordinate transformations in the Einstein frame target space~\cite{Sasaki:1995aw, Peterson:2010np, Gong:2011uw,Kaiser:2012ak}. The basis vector $\hat T^I$ is tangent to the inflationary trajectory, and $\hat N^I$ is  orthogonal to it:
\begin{align}
	\hat T^I = \frac{\dot \Phi^I}{\sqrt{G_{IJ} \dot \Phi^I \dot \Phi^J}}, 
	\quad \hat N^I = \frac{\omega^I}{\omega},
	\quad \omega^I = D_t \hat T^I,
\end{align}
where $\omega$ is the norm of the turn rate vector with
\begin{align}
	\omega = -\frac{V_{,N}}{\sqrt{G_{IJ} \dot \Phi^I \dot \Phi^J}}, 
	\qquad V_{,N} = \hat N^I \frac{\partial V}{\partial \Phi^I}.
\end{align}
Unit length and orthogonality imply
\begin{align}
	G_{IJ} \hat T^I \hat T^J =1 \, , \qquad  G_{IJ} \hat N^I \hat N^J =1, 
	\qquad G_{IJ} \hat N^I \hat T^J = 0.	
\end{align}
The components of $\hat T$ and $\hat N$ along the inflationary trajectory are given by
\begin{align}
	\hat T \simeq
	\begin{pmatrix} 1 \\ \sqrt{\frac{\xi}{6\left(4\lambda \alpha + \xi^2\right)}} e^{\frac{\sigma}{\sqrt{6}M_P} }
	\end{pmatrix}, 
	\qquad \hat N \simeq
	\pm
	\begin{pmatrix} 
	-\sqrt{\frac{\xi}{6\left(4\lambda \alpha + \xi^2\right)}} \\  e^{\frac{\sigma}{\sqrt{6}M_P}} 
	\end{pmatrix}.	
\end{align}
At first sight, it might look as if the second component of $\hat T$ is large,
but it is actually subdominant for $\xi \gg 1$ since it is always contracted with $G_{\sigma \sigma}$.
The turn rate inside the valley is then given by
\begin{align}
	\frac{\left\lvert \omega \right\rvert}{\mathcal{H}} \simeq \sqrt{\frac{3\xi}{2\left(4\lambda \alpha + \xi^2\right)}}
	\sim \mathcal{O}\left(\xi^{-1/2}\right)
	\ll 1,	
\end{align}
for $\xi \gg 1$,
confirming that the single-field approximation was reasonable. The smallness of the turn rate guarantees that isocurvature modes would not modify the curvature power spectrum~\cite{Kaiser:2012ak}. 

We can also show that the mass of the isocurvature mode is large,
again consistent with the valley approximation~\eqref{eq:valley}.
The isocurvature mass is found by projecting the mass matrix 
\begin{align}
	M_{IJ} = \nabla_I \nabla_J V + R_{IKJL} \dot \Phi^K \dot \Phi^L,
\end{align}
onto the $\hat{N}$-direction:
\begin{align}
	m_N^2 = \hat N^I \hat N^J \left(\nabla_I \nabla_J V + R_{IKJL} \dot \Phi^K \dot\Phi^L \right),
\end{align}
all evaluated at $\phi = \phi_v$. A straightforward computation shows that the ratio of the isocurvature mass over the Hubble parameter is
\begin{align}
	\frac{m_N^2}{3\mathcal H^2} \simeq - \frac{\epsilon_T}{9} - \sqrt\frac{\epsilon_T}{3} + \frac{2 \xi \left(4 \alpha \lambda + \xi^2\right)}{\alpha \lambda}\, ,
\end{align}
where $\epsilon_T$ is the  slow roll parameter defined along the direction $\hat T$
\begin{align}
	\epsilon_T = \frac{M_P^2}{2} \left( \frac{V_{,T}}{V} \right)^2\, ,
\end{align}
and in our case $\epsilon_T \sim \epsilon_V$ with $\epsilon_V$ defined in Eq.~\eqref{eq:sreps}. The only significant contribution to the isocurvature mass is the third term, implying that the isocurvature mass is positive and large compared to the Hubble parameter for $\xi \gg 1$. Thus the isocurvature power spectrum is suppressed, in agreement with observations~\cite{Akrami:2018odb}.

\section{Electroweak vacuum metastability}
\label{sec:quartic}
In Sec.~\ref{sec:inf}, we have assumed that the Higgs quartic coupling $\lambda$ is positive
at the inflationary scale. It is well-known, however, that $\lambda$ becomes negative
at the renormalization scale $\mu = 10^9\mathchar`-10^{10}\,\mathrm{GeV}$ 
for the current central value of the top quark mass within the SM~\cite{Degrassi:2012ry,Buttazzo:2013uya}.
This feature strongly depends on the top quark mass, and the absolute stability of the EW vacuum 
is still allowed given an uncertainty in the determination of the top quark mass.
In this section, we study how the scalaron affects this picture.

First of all, the scalaron cannot make $\lambda$ positive for all the energy scales
if we assume that inflation occurs within the Higgs-$R^2$ sector.
As we have seen in Sec.~\ref{sec:inf}, assuming that $\lambda > 0$,
the CMB normalization requires
\begin{align}
	\frac{\xi^2}{\lambda} + 4\alpha \simeq 2\times 10^9.
\end{align}
It implies that the scalaron mass is bounded from below as\footnote{
	Strictly speaking, we here assume that the Higgs mass term and the cosmological constant
	are negligible at the scalaron mass scale,
	which should be true to be consistent with the current universe.
}
\begin{align}
	m_\sigma = \frac{M_P}{2\sqrt{3\alpha}} \gtrsim 3\times 10^{13}\,\mathrm{GeV}.
\end{align}
The scalaron does not affect the RG flow of $\lambda$ below this scale.
Since $\lambda$ already turns negative below this scale
for the central value of the top quark mass,
the scalaron cannot make $\lambda > 0$ for all energy scales.
Nevertheless, the scalaron may affect the positivity of $\lambda$ above the scalaron mass scale,
which is the main topic of this section.

We separate our discussion into two parts. 
The first part focuses on the threshold corrections
that arise by matching parameters between low and high energy theories.
The second part is on the RG evolution of the parameters.
We note that the same topic is analyzed in Ref.~\cite{Gorbunov:2018llf}.
A difference is that 
we have included the full RGEs up to two-loop,
while (it seems that) they have included only the one-loop correction from $\xi$ and $\alpha$ to the running of $\lambda$.
Thus, our analysis here is more precise.

\subsection{Threshold correction}\label{sec:threshold}
In this subsection, we discuss the tree-level threshold correction.
Our conclusion is that there is no tree-level threshold correction that makes $\lambda$ positive.

First, it is important to notice that, although we refer to $\lambda$ as the Higgs quartic coupling,
it is not the coefficient of the Higgs quartic term above the scalaron mass scale, the UV theory. We refer to this tree-level quartic Higgs coupling as $\lambda_\mathrm{UV}$. It can be obtained from the potential in Eq.~\eqref{eq:action_conformal}
\begin{align}
	\lambda_\mathrm{UV} = \lambda + \frac{\left(\xi + 1/6\right)^2}{\alpha}.
	\label{eq:lambda_UV}
\end{align}
The tree-level matching at the scalaron mass scale is demonstrated in App.~B of Ref.~\cite{Ema:2017rqn}.\footnote{
	Ref.~\cite{Ema:2017rqn} ignores kinetic mixings among the scalar fields in the Einstein frame
	and hence the sub-leading terms suppressed by $\xi$,
	or the ``1/6" in Eq.~\eqref{eq:lambda_UV}.
}
There, $\lambda_m$ and $\lambda_\Lambda$ were not included, but this is consistent with the very small values of Eq.~\eqref{eq:bc_m_Lambda}, 
so we will also take $\lambda_m(\mu = m_\sigma)=\lambda_\Lambda(\mu=m_\sigma)=0$. 
The threshold correction is then obtained by comparing the 4-point interaction in the IR (which only includes the four-point vertex) and the UV (which includes a four-point vertex and diagrams with scalaron exchange).
One finds that the parameter $\lambda$ does not receive a threshold correction at tree level. 
This can also be seen from the fact that the $\lambda_\alpha$-term vanishes
by minimizing the scalar potential with respect to $\sigma$ for $\lambda_m = \lambda_\Lambda = 0$,
thus only the $\lambda$-term is left.
The EW vacuum stability is therefore not affected by the threshold correction, 
at least in the case $\xi^2 \sim \alpha \gg 1$.
Note that it is the positivity of $\lambda$, not $\lambda_\mathrm{UV}$, 
that is crucial for the Higgs to play (a part of) the inflaton in Higgs-$R^2$ inflation.

\subsection{RG evolution}

\begin{figure}[t]
	\centering
 	\includegraphics[width=0.49\linewidth]{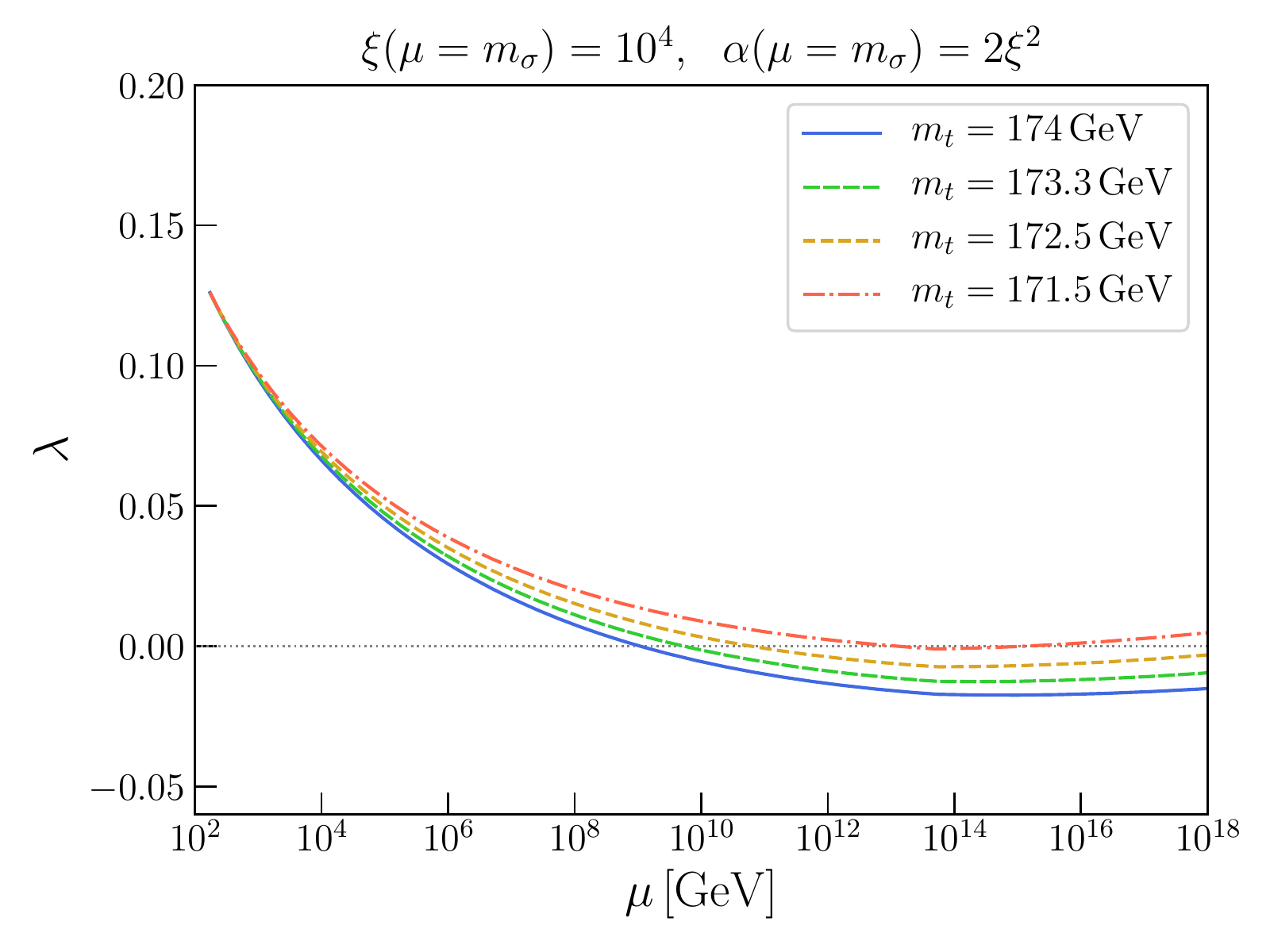}
 	\includegraphics[width=0.49\linewidth]{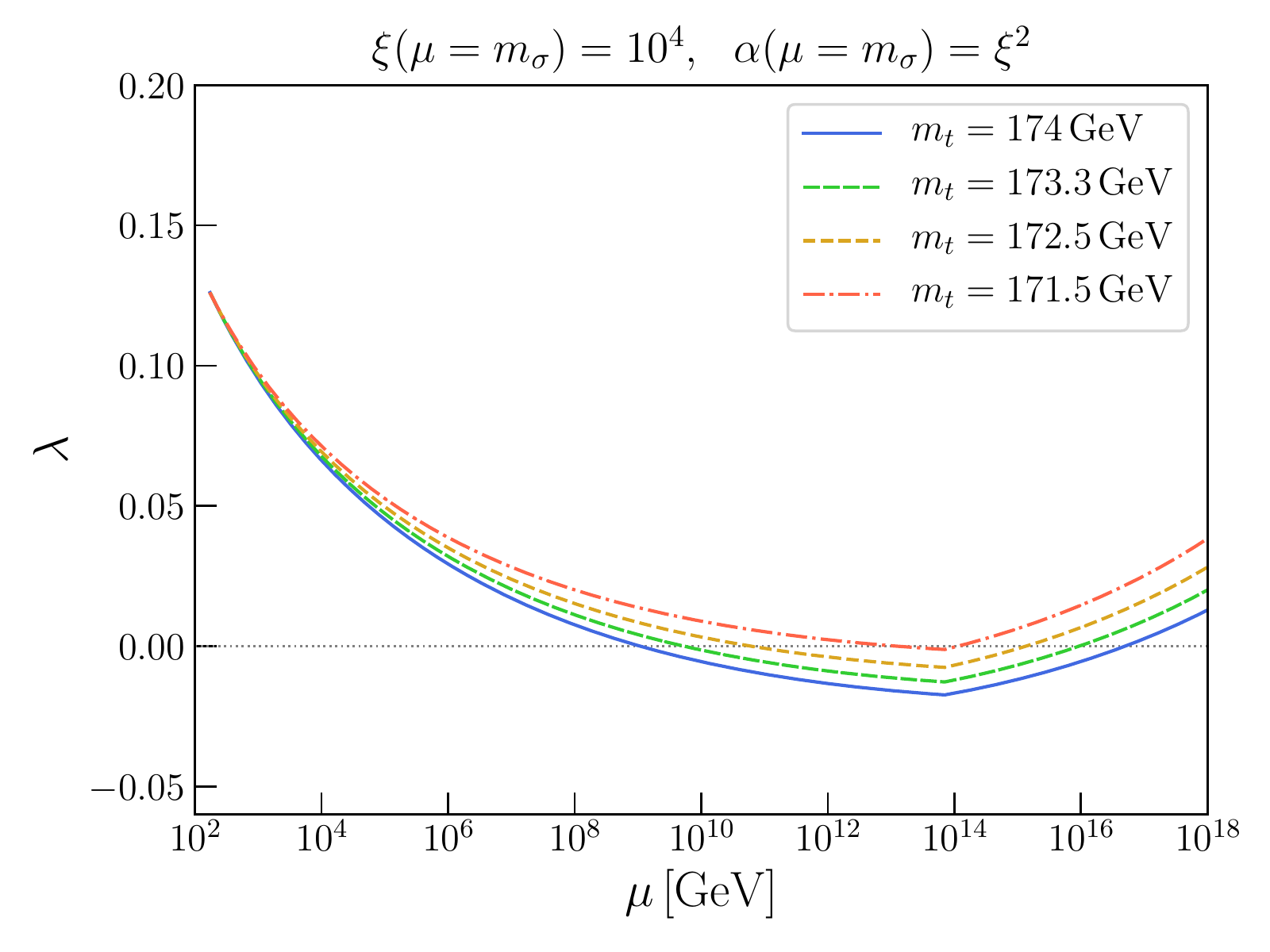}
 	\includegraphics[width=0.49\linewidth]{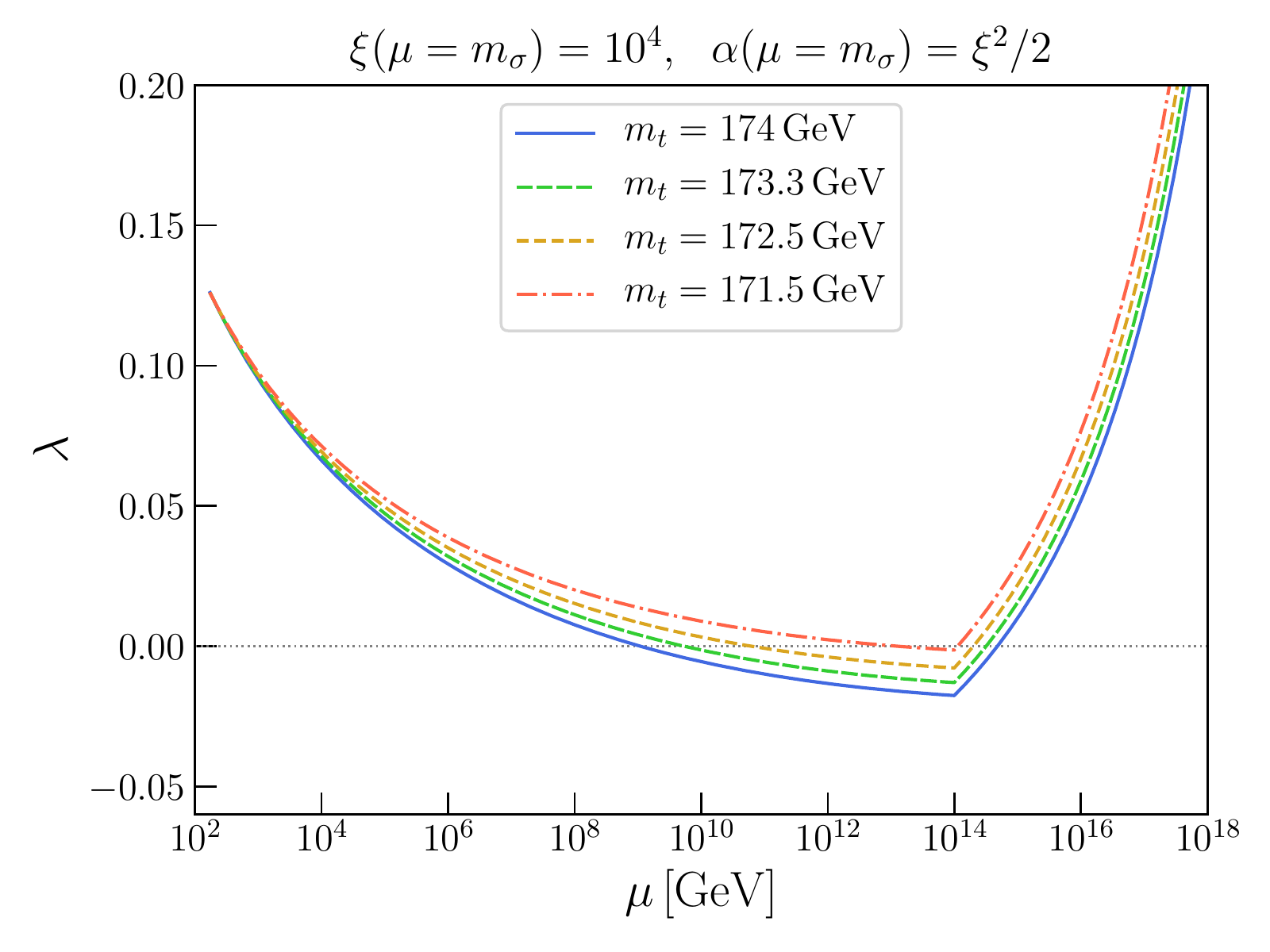}
 	\includegraphics[width=0.49\linewidth]{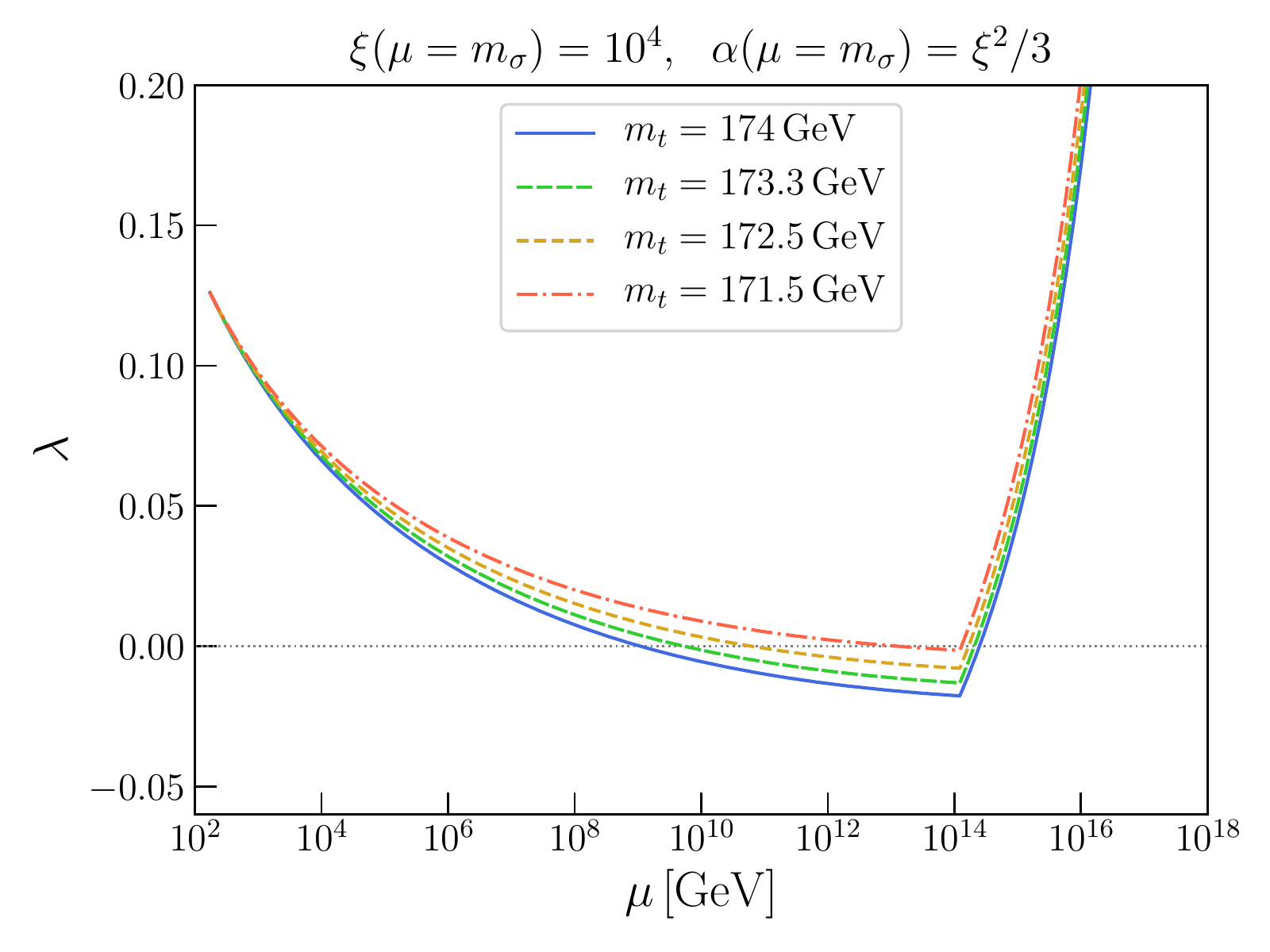}
	\caption{\small The RG running of $\lambda$ for different values of the ratio $\xi^2/\alpha$
	with $\xi (\mu = m_\sigma) = 10^4$.
	The additional contribution from the scalaron makes $\lambda$ positive
	if $\xi^2/\alpha$ is large enough.
	A too large value of $\xi^2/\alpha$ however results in a Landau pole below the Planck scale.
	The Higgs mass is fixed as $m_h = 125.15\,\mathrm{GeV}$, and the other SM parameters
	are fixed according to Ref.~\cite{Buttazzo:2013uya}.
	We have used the two-loop SM beta functions for $m_t < \mu < m_\sigma$,
	and the beta functions of Secs.~\ref{subsec:one_loop} and~\ref{subsec:two_loop} 
	for $m_\sigma < \mu < M_P$.}
	\label{fig:rge_lambda}
\end{figure}

Although the threshold correction does not affect $\lambda$, the scalaron can make $\lambda$ positive
through the RG evolution, as pointed out in Ref.~\cite{Gorbunov:2018llf}.
The reason is that scalar fields in general give positive contributions to the beta function of $\lambda$.
In Fig.~\ref{fig:rge_lambda}, we show the RG running of $\lambda$ 
for several parameter sets.
We have used the two-loop SM RGEs below the scalaron mass scale $m_\sigma = M_P/12\alpha$,
and the full RGEs up to two-loop given in Secs.~\ref{subsec:one_loop} and~\ref{subsec:two_loop}
above $m_\sigma$.
We have used, however, only the tree-level matching conditions at $m_\sigma$,
and hence Fig.~\ref{fig:rge_lambda} should be understood as a demonstration,
not having the full two-loop precision.
It is beyond the scope of this paper to compute the one-loop matching conditions,
and we leave it for future work.
The boundary conditions for the Higgs mass and the cosmological constant
are $\lambda_m(\mu = m_\sigma) = \lambda_\Lambda(\mu = m_\sigma) = 0$.
We have fixed $\xi = 10^4$ and varied the ratio $\xi^2/\alpha$ 
as $\xi^2/\alpha = 1/2, 1, 2, 3$
at $\mu = m_\sigma$
since the additional contribution to $\beta_\lambda$ is controlled by this ratio.
One can see that $\lambda$ turns positive at high energy
if the ratio $\xi^2/\alpha$ is large enough.
A too large $\xi^2/\alpha$ however makes $\lambda$ too large 
so that the theory hits a Landau pole and loses perturbativity below the Planck scale.

As we have seen, 
the scalaron cannot make $\lambda$ positive for all energy scales,
yet it can push $\lambda$ to be positive at high energy for a relatively large value of $\xi^2/\alpha$.
As a result, the scalar potential may develop a local minimum around $m_\sigma$.
It would be interesting to discuss the cosmological implication of this local minimum,
especially during and after inflation.
Cosmological implications of EW vacuum metastability have been studied in detail in the literature.
For instance, if the inflationary scale is too high, it triggers EW vacuum decay during inflation,
resulting in an upper bound on the inflationary scale~\cite{Espinosa:2007qp,Lebedev:2012sy,Kobakhidze:2013tn,Fairbairn:2014zia,Hook:2014uia,Kamada:2014ufa,Herranen:2014cua,Kearney:2015vba,Espinosa:2015qea,East:2016anr,Joti:2017fwe,Rajantie:2017ajw,Fumagalli:2019ohr}.
In addition, the inflaton-Higgs couplings or the non-minimal coupling between the Higgs and the Ricci scalar
can produce a large amount of Higgs quanta during the preheating epoch, causing EW vacuum decay after inflation
and hence putting upper bounds on these couplings~\cite{Herranen:2015ima,Ema:2016kpf,Kohri:2016wof,Enqvist:2016mqj,Postma:2017hbk,Ema:2017loe,Ema:2017rkk,Figueroa:2017slm,Rusak:2018kel,Croon:2019dfw}.
These works usually assume that the Higgs potential is negative (at least) up to the Planck scale,
and hence it is difficult to recover the EW vacuum once the Higgs field rolls down to the negative region.
In the current case, the situation can be different, 
as the Higgs potential may be again positive at high energy.
Hence, even if the Higgs is trapped at the local minimum at some epoch in the early universe,
it may be easier to restore the EW vacuum by, \textit{e.g.}, thermal effects.
We leave a detailed study for future work.

\section{Conclusion and discussion}
\label{sec:conclusion}
Higgs-$R^2$ inflation is an interesting model, as it gives inflationary predictions consistent with the Planck data~\cite{Akrami:2018odb}. It includes a new scalaron degree of freedom at the inflationary scale
on top of Higgs inflation, which makes the theory unitary and renormalizable up to the Planck scale (in contrast with Higgs inflation without a scalaron). In the present work, we have presented the one- and two-loop Renormalization Group Equations for this system and studied their phenomenological consequences. The RGEs are valid up to the Planck scale in the large $\xi$ regime that we studied (which is the most relevant for inflation). 
Below we summarize each section of this paper and list possible future directions.

\subsubsection*{One- and two-loop RGEs}
In Sec.~\ref{subsec:weyl},
to derive the RGEs, we decomposed the metric into the determinant part, which we call the conformal mode, and a spin-2 part. We show in Sec.~\ref{subsec:gauge}
that the spin-2 part couples to the matter fields only via Planck-suppressed operators and we can thus neglect $h_{\mu\nu}$ below the Planck scale.
Furthermore, for our gauge fixing condition, Eq.~\eqref{eq:gauge_fix}, 
we can also ignore the Faddeev-Poppov ghost field. 
 These properties greatly simplify our computations.

We thus derived the RGEs from Eq.~\eqref{eq:LSM}, which contains only the conformal mode, the scalaron, the Higgs field and the other SM fields. The one-loop RGEs are presented in Sec.~\ref{subsec:one_loop}. These were obtained by computing four-point functions explicitly. We cross-checked the RGEs with the help of \texttt{SARAH} and confirmed that the RGEs match the results of Refs.~\cite{Salvio:2014soa,Salvio:2017qkx,Salvio:2018crh} if we ignore the spin-2 contribution. The details of the computation are given in App.~\ref{app:one_loop}. 
Using \texttt{SARAH}, we also obtained the two-loop RGEs in Sec.~\ref{subsec:two_loop}. 
App.~\ref{app:two_loop} contains the computational details. To our knowledge, the two-loop RGEs of 
the Higgs-$R^2$ theory are presented for the first time in this present work. 

\subsubsection*{Consequences for inflation and reheating}
The interactions corresponding to the couplings $\lambda_m$, which generates a Higgs mass, and $\lambda_\Lambda$, the cosmological constant, have to be included to renormalize the theory. These terms were not taken into account in previous studies of the Higgs-$R^2$-model. 
Therefore, in Sec.~\ref{sec:inf}, we analytically study the inflationary prediction with a nonzero $\lambda_m$ and $\lambda_\Lambda$. The predictions for the single-field inflationary trajectory along the valley are not affected by the nonzero values of $m$ and $\Lambda$, and are thus consistent with the Planck data. Although the inflationary predictions are unaffected, the presence of the mass term and the cosmological constant might affect the particle production stage at the end of inflation. Reheating in the Higgs-$R^2$ system was studied in Refs.~\cite{He:2018mgb, Bezrukov:2019ylq, He:2020ivk, Bezrukov:2020txg}, but these studies did not include nonzero $m$ and $\Lambda$. In Ref.~\cite{Fu:2019qqe} reheating was studied in a similar model, with a nonzero mass term for the field that resembles our Higgs field. The study suggests that the nonzero mass term leads to 
a smaller variance of the Higgs-like fields in the rescattering regime and also affects the evolution of the equation of state. 

\subsubsection*{Electroweak stability}
In Sec.~\ref{sec:quartic} we studied another phenomenological consequence of the RGEs: the running of $\lambda$, which is relevant for the stability of the Higgs potential. The inclusion of the scalaron can not make $\lambda$ positive for all energy scales, since the scalaron mass scale is above the scale where $\lambda$ typically runs negative (this scale depends on the value of the top quark mass),
assuming that inflation happens in the Higgs-$R^2$ sector. 
Nevertheless, the Higgs potential can be stabilized at larger energy scales. This stabilization mechanism depends on the ratio $\xi^2/\alpha$. If the ratio is too small, the stabilizing effect is also small. However, if the ratio is too large, $\lambda$ encounters a Landau pole below the Planck scale. Let us point out that we used only tree-level matching at the scalaron mass scale in this paper.
In this case $\lambda$ receives no stabilizing threshold correction. We leave the computation of the threshold correction in a one-loop matching procedure for future work.
\subsubsection*{Residual gauge symmetry}
The fact that there are only six independent parameters in Eq.~\eqref{eq:LSM} and the theory can thus be renormalized by only six counterterms, instead of possibly nine, suggests that the shape of the potential is restricted by some symmetry. In App.~\ref{sec:renormalizability} we show that the requirement that the action is invariant under (the scalar part of) the residual gauge symmetry $x^\mu \rightarrow x^\mu - \partial^\mu \xi$, restricts the allowed interactions. The residual gauge symmetry can explain the shape of the potential, but only if we do not include the scalaron from the beginning, but obtain it from the $\left( \Box \Phi/\Phi\right) ^2$-term. To us, this approach seems somewhat ad hoc and a better understanding is lacking. The physical reason for including the operators $\left( \Box \Phi/\Phi \right)^2$ and $|H|^2 \Box \Phi/\Phi$, but neglecting all other higher derivative operators is also not clear. We leave a better understanding of these points for future work. 

\subsubsection*{Generalization to other theories/asymptotic safety}
In this paper we have focused on the Higgs-$R^2$ theory, \textit{i.e.}, 
the theory with the $R^2$-term and the SM.
Our method can be, however, applied to more general theories
which contain an $R^2$-term and arbitrary numbers of fields with
generic Yukawa, gauge and scalar quartic interactions.
Indeed, it is expected to be straightforward to obtain the RGEs of such a general theory
with the help of, \textit{e.g.}, 
Refs.~\cite{Machacek:1983tz,Machacek:1983fi,Machacek:1984zw,Luo:2002ti} and the procedure outlined in App.~\ref{app:two_loop}.
These RGEs may be useful beyond the context of inflation,
\textit{e.g.}, for the asymptotic safety program~\cite{Weinberg:1976xy,Weinberg:1980gg}
(see Ref.~\cite{Eichhorn:2018yfc} for a recent review).

An ultimate goal of asymptotic safety is to understand the UV structure of quantum gravity.
Our method cannot be directly connected to this ultimate goal since we ignore the spin-2 particles.
Our method also assumes that the theory is perturbative, and is useless in a strong coupling regime.
Nevertheless it correctly captures the effect of the scalar sector of the gravity,
and hence may serve as a small step toward the ultimate goal of the asymptotic safety program.
For instance,
a UV fixed point is guaranteed to exist in a general renormalizable theory without gravity
in a perturbative regime in the Veneziano limit~\cite{Litim:2014uca}.
We can see how this conclusion is affected
by an inclusion of the $R^2$-term and hence the conformal mode of the metric and the scalaron.
It might also be interesting to study the LSM~\eqref{eq:LSM}
by the functional RG approach~\cite{Wilson:1973jj,Wetterich:1992yh,Morris:1993qb,Reuter:1996cp}
since it allows us to go beyond the perturbative regime.\footnote{
	The functional RG approach is also applied to Higgs inflation~\cite{Saltas:2015vsc} and $R^2$ inflation~\cite{Copeland:2013vva}.
}~\footnote{
	We believe that our method is similar in spirit as Refs.~\cite{Lauscher:2002sq,Machado:2007ea,Codello:2008vh,Falls:2013bv,Falls:2014tra}
	since they only focus on the $R^n$-terms that 
	affect the scalar sector of the theory.
}
Finally, it might be interesting by itself that
the LSM~\eqref{eq:LSM} without any matter fields has a fixed point
$\lambda_\alpha = 0$ for an arbitrary value of $\lambda_\Lambda$,
which is also discussed in Ref.~\cite{Copeland:2013vva}.

\section*{Acknowledgement}
This work was funded by the Deutsche Forschungsgemeinschaft under Germany’s Excellence
Strategy - EXC 2121 “Quantum Universe” - 390833306.
The Feynman diagrams are drawn by \texttt{TikZ-Feynman}~\cite{Ellis:2016jkw}. 
 
\appendix

\section{General coordinate transformation}
\label{app:general_coord_transf}
In this appendix, we derive Eqs.~\eqref{eq:varphi_transf} and~\eqref{eq:tensor_transf}.
Under the general coordinate transformation, $x^\mu \rightarrow x^\mu - \xi^\mu$, 
the spacetime metric transforms as
\begin{align}
	g_{\mu\nu} \rightarrow
	g'_{\mu\nu} 
	= g_{\mu\nu} + \left(\partial_\mu \xi^\alpha\right)g_{\alpha \nu} + g_{\mu \alpha}\left(\partial_\nu \xi^\alpha\right)
	+ \xi^\alpha\partial_\alpha g_{\mu\nu},
\end{align}
up to the first order in $\xi^\mu$.
By substituting the decomposition~\eqref{eq:metric_decomposition}, it reads
\begin{align}
	e^{2\varphi'}[e^{h'}]_{\mu\nu}
	&= 
	e^{2\varphi}\left\{[e^{h}]_{\mu\nu} 
	+ \left(\partial_\mu \xi^\alpha\right)[e^{h}]_{\alpha\nu}
	+ [e^{h}]_{\mu\alpha}\left(\partial_\nu \xi^\alpha\right)
	+ 2\xi^\alpha \left(\partial_\alpha \varphi\right)[e^h]_{\mu\nu} 
	+ \xi^\alpha \partial_\alpha [e^{h}]_{\mu\nu},
	\right\} 
\end{align}
where the indices are raised and lowered by the flat spacetime metric $\eta_{\mu\nu}$ here and hereafter.
It can be rewritten as
\begin{align}
	e^{2\left(\varphi' - \varphi\right)}[e^{h'}]_{\mu\nu}
	= {[e^{h/2}]_\mu}^\rho
	&\left\{
	\left(1+2\xi^\alpha \partial_\alpha \varphi\right)\delta_{\rho}^{\sigma}
	+ {[e^{-h/2}]_{\rho}}^{\beta} \left(\partial_\beta \xi^\alpha\right){[e^{h/2}]_{\alpha}}^{\sigma}
	\right. \nonumber \\ &\left.
	+ [e^{h/2}]_{\rho\alpha} \left(\partial_\beta \xi^\alpha\right)[e^{-h/2}]^{\beta\sigma}	
	+ {[e^{-h/2}]_\rho}^\beta \left(\xi^\alpha \partial_\alpha [e^{h}]_{\beta\gamma}\right)[e^{-h/2}]^{\gamma\sigma}
	\right\}
	[e^{h/2}]_{\sigma\nu}.
	\label{eq:full_transf}
\end{align}
By taking the determinant of both sides, we obtain
\begin{align}
	e^{8\left(\varphi'-\varphi\right)}
	= 1 + 8\xi^\alpha \partial_\alpha \varphi + 2 \partial_\alpha \xi^\alpha 
	+ \xi^\alpha \mathrm{tr}\left[e^{-h}\partial_\alpha e^{h} \right]
	+ \mathcal{O}(\xi^2),
\end{align}
where we have used the relation
\begin{align}
	\mathrm{Det}\left[\delta + \epsilon A\right] = 1 + \epsilon\,\mathrm{Tr}A + \mathcal{O}(\epsilon^2),
\end{align}
with $\delta^\nu_\mu$ being the identity matrix.
We can use Jacobi's formula to show that
\begin{align}
	\mathrm{Tr}\left[e^{-h}\partial_\alpha e^{h}\right]
	= \partial_\alpha \ln\mathrm{Det}\left[e^{h}\right] = 0,
\end{align}
where we have used the definition of the decomposition~\eqref{eq:metric_decomposition} in the second equality.
Thus, we obtain the transformation law of the conformal mode of the metric as
\begin{align}
	\varphi' = \varphi + \xi^\alpha \partial_\alpha \varphi + \frac{1}{4} \partial_\alpha \xi^\alpha,
\end{align}
up to first order in $\xi^\mu$.
By substituting into Eq.~\eqref{eq:full_transf}, we obtain
\begin{align}
	\left[e^{-h/2}e^{h'}e^{-h/2}\right]_{\mu\nu} =& 
	\left(1-\frac{1}{2}\left(\partial_\alpha \xi^\alpha\right)\right)\eta_{\mu\nu}
	+ \left[e^{-h/2} \left(\partial \xi\right) e^{h/2}\right]_{\mu\nu} + \left[e^{-h/2}\left(\partial \xi\right) e^{h/2}\right]_{\nu\mu} 
	+ \left[e^{-h/2}\left(\xi^\alpha\partial_\alpha e^{h}\right)e^{-h/2}\right]_{\mu\nu},
	\label{eq:tensor_transf_med}
\end{align}
where we have relied on the matrix notation, and $[\partial \xi]_{\mu\nu} \equiv \partial_\mu \xi_\nu$.
We now simplify this equation further.
For this purpose, it is useful to define the adjoint action as
\begin{align}
	\mathrm{ad}_{X} Y \equiv \left[X, Y\right] = XY - YX,
\end{align}
for arbitrary matrices $X$ and $Y$.
Then the following relation holds:
\begin{align}
	e^{X} Y e^{-X} = e^{\mathrm{ad}_X} Y,
\end{align}
where the right-hand-side should be understood by the Taylor expansion as
\begin{align}
	e^{\mathrm{ad}_X} Y = Y + \left[X, Y\right] + \frac{1}{2!}\left[X, \left[X,Y\right]\right]
	+ \frac{1}{3!}\left[X, \left[X, \left[X, Y\right]\right]\right] + \cdots.
\end{align}
One can easily see that the adjoint action is linear, \textit{i.e.},
\begin{align}
	\mathrm{ad}_{cX} = c\, \mathrm{ad}_X,
	\quad
	\mathrm{ad}_{X+Y} = \mathrm{ad}_X + \mathrm{ad}_Y,
\end{align}
for an arbitrary complex number $c$.
The variation of an exponentiated matrix is expressed with the help of the adjoint action as
\begin{align}
	e^{X + \Delta X} - e^X = e^{X}\frac{1-e^{-\mathrm{ad}_X}}{\mathrm{ad}_X}\Delta X.
\end{align}
Equipped with these expressions, Eq.~\eqref{eq:tensor_transf_med} is simplified as
\begin{align}
	\left[\left(\frac{e^{\mathrm{ad}_{h}/2}-e^{-\mathrm{ad}_h/2}}{\mathrm{ad}_h}\right)\left(h'-h\right)\right]_{\mu\nu}
	=& -\frac{1}{2}\left(\partial_\alpha\xi^\alpha\right)\eta_{\mu\nu}
	+ \left[e^{-\mathrm{ad}_h/2} \partial \xi\right]_{\mu\nu}
	+ \left[e^{-\mathrm{ad}_h/2} \partial \xi\right]_{\nu\mu}
	+ \left[\left(\frac{e^{\mathrm{ad}_{h}/2}-e^{-\mathrm{ad}_h/2}}{\mathrm{ad}_h}\right)\left(\xi^\alpha \partial_\alpha h\right)\right]_{\mu\nu},
	\label{eq:tensor_transf_before_inverse}
\end{align}
to first order in $\xi^\mu$.
The adjoint action for a symmetric tensor $X$ (\textit{i.e.} $X^\mathrm{T} = X$) satisfies
\begin{align}
	\mathrm{ad}_X Y^\mathrm{T} = - \left[\mathrm{ad}_X Y\right]^\mathrm{T},
\end{align}
where the superscript $\mathrm{T}$ indicates the transpose.
Therefore, by acting with the inverse operator from the left on Eq.~\eqref{eq:tensor_transf_before_inverse}, 
we finally obtain the transformation law of $h_{\mu\nu}$ as
\begin{align}
	h'_{\mu\nu} = h_{\mu\nu}
	+ \left[\left(\frac{\mathrm{ad}_h}{e^{\mathrm{ad}_h} -1}\right) \partial \xi\right]_{\mu\nu}
	+ \left[\left(\frac{\mathrm{ad}_h}{e^{\mathrm{ad}_h} -1}\right) \partial \xi\right]_{\nu\mu}
	- \frac{1}{2}\left(\partial_\alpha \xi^\alpha\right)\eta_{\mu\nu}
	+ \xi^\alpha \partial_\alpha h_{\mu\nu},
\end{align}
up to first order in $\xi^\mu$.
It completes our derivation of Eqs.~\eqref{eq:varphi_transf} and~\eqref{eq:tensor_transf}.
Note that the traceless property of $h_{\mu\nu}$ is indeed preserved since
\begin{align}
	\mathrm{Tr}\left[\left(\frac{\mathrm{ad}_h}{e^{\mathrm{ad}_h} -1}\right) \partial \xi\right]
	= \partial_\alpha \xi^\alpha.
\end{align}

\section{Einstein and conformal frame}
\label{app:eins-conf}
In this appendix, we clarify the relation between the action given in Eq.~\eqref{eq:action_conformal} and that in the Einstein frame \eqref{eq:action_einstein}.
Let us start from Eq.~\eqref{eq:action_einstein}.
The metric in the Einstein frame can be decomposed in the same way
\begin{align}
	g_{\mu\nu} = e^{2\varphi_E} \tilde g_{\mu\nu}, \quad 
	\operatorname{Det}\left[\tilde{g}_{\mu\nu}\right] = -1.
\end{align}
The conformal mode in the Einstein frame is defined as
\begin{align}
	\Phi_E = \sqrt{6} M_P e^{\varphi_E}.
\end{align}
Here we use the subscript $E$ to clarify that $\Phi_E$ is different from $\Phi$.
We also rescale the fields as
\begin{align}
	\sigma_E \to e^{ - \varphi_E } \sigma_E, \quad 
	H \to e^{\sigma_E/\Phi_E - \varphi_E } H, \quad
	\psi \to e^{- 3 \varphi_E /2} \psi.
\end{align}
The scalaron field $\sigma$ given in Eq.~\eqref{eq:action_conformal} is different from $\sigma_E$.
The relation between them is given by
\begin{align}
	\frac{\sigma_E}{\Phi_E} = - \ln \left( \sqrt{\Phi_E^2 + 2 \abs{H}^2 + \sigma^2} + \sigma \right) + \ln \Phi_E.
\end{align}
Inserting these equations into Eq.~\eqref{eq:action_einstein}, we obtain
\begin{align}
	S = \int \dd^4 x &\Bigg[
		\frac{\tilde R}{12} \Phi_E^2
		+ \tilde g^{\mu\nu} D_\mu H^\dag D_\nu H
		+ \frac{1}{2} \tilde g^{\mu\nu} \partial_\mu \sigma \partial_\nu \sigma
		\nonumber \\
		& - \frac{1}{2} \tilde g^{\mu\nu} \partial_\mu \sqrt{\Phi_E^2 + 2 \abs{H}^2 + \sigma^2} \partial_\nu \sqrt{\Phi_E^2 + 2 \abs{H}^2 + \sigma^2} - V \left( \sqrt{\Phi_E^2 + 2 \abs{H}^2 + \sigma^2}, H, \sigma \right)
	\Bigg] + S_{\psi + A} \left[ \tilde g_{\mu\nu} \right],
\end{align}
where
\begin{align}
	V (\Phi, H ,\sigma) = 
	\lambda \abs{H}^4 + \frac{\lambda_m}{2} \left( \Phi + \sigma \right)^2 \abs{H}^2
		+ \frac{\lambda_\Lambda}{4} \left( \Phi + \sigma \right)^4
		+ \frac{\lambda_\alpha}{4} \left[ \sigma \left( \Phi + \sigma \right) + 2 \bar\xi \abs{H}^2 \right]^2.
\end{align}
Therefore we reproduce Eq.~\eqref{eq:action_conformal} by the following relation between $\Phi_E$ and $\Phi$:
\begin{align}
	\Phi_E^2 = \Phi^2 - 2 \abs{H}^2 - \sigma^2.
\end{align}

\section{Computational details of one-loop RGEs}
\label{app:one_loop}
In this appendix, we provide some intermediate steps to derive the one-loop RGEs.
It is shown in the main text that the SM fermions and gauge bosons do not couple 
to the conformal mode $\Phi$ nor the scalaron $\sigma$ thanks to the Weyl invariance.
Therefore, their effects on the RGEs except for the Higgs quartic coupling $\lambda$
are only through the Higgs anomalous dimension, which is easy to reproduce.
For this reason, we ignore the SM fermions and gauge bosons and 
focus on the scalar sector of the Higgs-$R^2$ theory in this appendix.

\subsection{Feynman rules}

As we have explained the main text, 
we can simply take $\tilde{g}_{\mu\nu} = \eta_{\mu\nu}$ without
worrying about any complications associated with the ghost fields below $M_P$.
Thus, the action from which we compute the RGEs is
\begin{align}
	S
	= \int \dd^4 x &\left[
	- \frac{1}{2}\eta^{\mu\nu} \partial_\mu \Phi \partial_\nu \Phi
	+ \frac{1}{2}\eta^{\mu\nu}\partial_\mu \phi_i \partial_\nu \phi_i
	+ \frac{1}{2}\eta^{\mu\nu}\partial_\mu \sigma \partial_\nu \sigma
	- \frac{\lambda}{4} \phi_i^4 
	\right. \nonumber \\ &\left.
	- \frac{\lambda_m}{4}\left(\Phi + \sigma \right)^2\phi_i^2
	- \frac{\lambda_\Lambda}{4}\left(\Phi+\sigma\right)^4
	- \frac{\lambda_\alpha}{4}\left(\sigma \left(\Phi + \sigma\right)
	+ \bar{\xi} \phi_i^2 \right)^2
	\right],
\end{align}
where we have slightly generalized the action by allowing the index $i$ to run from $i = 1$ to $i= N$,
where $N$ is the number of the real scalar fields ($N=4$ for the SM Higgs).
Note that we do not even have to expand $\Phi$ around $\sqrt{6}M_P$ to compute the RGEs.
The Feynman rules are readily derived from this action.
The propagators are given by
\begin{align}
	\Phi:~
	&\begin{tikzpicture}[baseline=(a)]
	\begin{feynman}[inline = (base.a), horizontal=a to c]
		\vertex (a);
		\vertex [above = 0.2em of a] (b);
		\vertex [right = of b] (c);
		\diagram*{
		(b) -- [photon] (c),
		};
	\end{feynman}
	\end{tikzpicture}
	= - \frac{i}{q^2}, \\
	\sigma:~
	&\begin{tikzpicture}[baseline=(a)]
	\begin{feynman}[inline = (base.a), horizontal=a to c]
		\vertex (a);
		\vertex [above = 0.2em of a] (b);
		\vertex [right = of b] (c);
		\diagram*{
		(b) -- (c),
		};
	\end{feynman}
	\end{tikzpicture}
	= \frac{i}{q^2}, \\
	\phi_i:~	
	&\begin{tikzpicture}[baseline=(a)]
	\begin{feynman}[inline = (base.a), horizontal=a to c]
		\vertex (a);
		\vertex [label=\(i\), above = 0.2em of a] (b);
		\vertex [label=\(j\),right = of b] (c);
		\diagram*{
		(b) -- [scalar] (c),
		};
	\end{feynman}
	\end{tikzpicture}
	= \frac{i}{q^2}\delta_{ij}.
\end{align}
where we have used the wavy line for $\Phi$ and the solid line for $\sigma$
since we do not consider the fermions and gauge bosons in this appendix.
They should not be confused with the SM fermions and gauge bosons.
Here it is important to notice the additional minus sign in the propagator of the conformal mode $\Phi$.
It originates from the ghost-like property of $\Phi$. 
Although ghost-like, it does not spoil the theory thanks to a residual gauge symmetry, 
as discussed in detail in Ref.~\cite{Ema:2020zvg}.
The Feynman rules for the interactions are given by
\begin{alignat}{2}
	\begin{tikzpicture}[baseline=(c)]
		\begin{feynman}[inline = (base.c)]
		\vertex (c);
		\vertex [above left = of c] (a);
		\vertex [below left = of c] (b);
		\vertex [above right = of c] (d);
		\vertex [below right = of c] (e);
		\diagram*{
		(a) -- (c),
		(b) -- (c),
		(d) -- (c),
		(e) -- (c),
		};
	\end{feynman}
	\end{tikzpicture}
	&= -6i\left(\lambda_\alpha + \lambda_\Lambda\right),
	\qquad
	&&\begin{tikzpicture}[baseline=(c)]
		\begin{feynman}[inline = (base.c)]
		\vertex (c);
		\vertex [above left = of c] (a);
		\vertex [below left = of c] (b);
		\vertex [above right = of c] (d);
		\vertex [below right = of c] (e);
		\diagram*{
		(a) -- (c),
		(b) -- (c),
		(d) -- (c),
		(e) -- [photon] (c),
		};
	\end{feynman}
	\end{tikzpicture}
	= -3i\left(\lambda_\alpha + 2\lambda_\Lambda\right), \\
	\begin{tikzpicture}[baseline=(c)]
		\begin{feynman}[inline = (base.c)]
		\vertex (c);
		\vertex [above left = of c] (a);
		\vertex [below left = of c] (b);
		\vertex [above right = of c] (d);
		\vertex [below right = of c] (e);
		\diagram*{
		(a) -- (c),
		(b) -- (c),
		(d) -- [photon] (c),
		(e) -- [photon] (c),
		};
	\end{feynman}
	\end{tikzpicture}
	&= -i\left(\lambda_\alpha + 6\lambda_\Lambda\right),
	\qquad
	&&\begin{tikzpicture}[baseline=(c)]
		\begin{feynman}[inline = (base.c)]
		\vertex (c);
		\vertex [above left = of c] (a);
		\vertex [below left = of c] (b);
		\vertex [above right = of c] (d);
		\vertex [below right = of c] (e);
		\diagram*{
		(a) -- (c),
		(b) -- [photon] (c),
		(d) -- [photon] (c),
		(e) -- [photon] (c),
		};
	\end{feynman}
	\end{tikzpicture}
	= -6i\lambda_\Lambda,
	\qquad
	\begin{tikzpicture}[baseline=(c)]
		\begin{feynman}[inline = (base.c)]
		\vertex (c);
		\vertex [above left = of c] (a);
		\vertex [below left = of c] (b);
		\vertex [above right = of c] (d);
		\vertex [below right = of c] (e);
		\diagram*{
		(a) -- [photon] (c),
		(b) -- [photon] (c),
		(d) -- [photon](c),
		(e) -- [photon] (c),
		};
	\end{feynman}
	\end{tikzpicture}
	= -6i \lambda_\Lambda, \\
	\begin{tikzpicture}[baseline=(c)]
		\begin{feynman}[inline = (base.c)]
		\vertex (c);
		\vertex [label=\(i\), above left = of c] (a);
		\vertex [label=270:\(j\), below left = of c] (b);
		\vertex [above right = of c] (d);
		\vertex [below right = of c] (e);
		\diagram*{
		(a) -- [scalar] (c),
		(b) -- [scalar] (c),
		(d) -- (c),
		(e) -- (c),
		};
	\end{feynman}
	\end{tikzpicture}
	&= -i\left(2\bar{\xi}\lambda_\alpha + \lambda_m\right) \delta_{ij},
	\qquad
	&&\begin{tikzpicture}[baseline=(c)]
		\begin{feynman}[inline = (base.c)]
		\vertex (c);
		\vertex [label=\(i\), above left = of c] (a);
		\vertex [label=270:\(j\), below left = of c] (b);
		\vertex [above right = of c] (d);
		\vertex [below right = of c] (e);
		\diagram*{
		(a) -- [scalar] (c),
		(b) -- [scalar] (c),
		(d) -- (c),
		(e) -- [photon] (c),
		};
	\end{feynman}
	\end{tikzpicture}
	= -i\left(\bar{\xi}\lambda_\alpha + \lambda_m\right) \delta_{ij}, \\
	\begin{tikzpicture}[baseline=(c)]
		\begin{feynman}[inline = (base.c)]
		\vertex (c);
		\vertex [label=\(i\), above left = of c] (a);
		\vertex [label=270:\(j\), below left = of c] (b);
		\vertex [above right = of c] (d);
		\vertex [below right = of c] (e);
		\diagram*{
		(a) -- [scalar] (c),
		(b) -- [scalar] (c),
		(d) -- [photon] (c),
		(e) -- [photon] (c),
		};
	\end{feynman}
	\end{tikzpicture}
	&= -i \lambda_m \delta_{ij},
	\qquad
	&&\begin{tikzpicture}[baseline=(c)]
		\begin{feynman}[inline = (base.c)]
		\vertex (c);
		\vertex [label=\(i\), above left = of c] (a);
		\vertex [label=270:\(j\), below left = of c] (b);
		\vertex [label=\(k\), above right = of c] (d);
		\vertex [label=270:\(l\), below right = of c] (e);
		\diagram*{
		(a) -- [scalar] (c),
		(b) -- [scalar] (c),
		(d) -- [scalar] (c),
		(e) -- [scalar] (c),
		};
	\end{feynman}
	\end{tikzpicture}
	= -2i\left(\bar{\xi}^2\lambda_\alpha + \lambda\right) \left(\delta_{ij}\delta_{kl} + \delta_{ik}\delta_{jl} + \delta_{il}\delta_{jk}\right),
\end{alignat}
where the combinatory factors from the external states are taken into account.

\subsection{Divergent part of four-point functions}
We compute the scalar four-point functions at one-loop level.
Below we list the divergent parts of the four-point functions before renormalization,
where we have performed dimensional regularization with $d = 4 - 2\epsilon$.
The divergent parts are
\begin{align}
	\left. iA_{\sigma^4}\right\rvert_{\mathrm{div.}} &= \frac{3i}{32\pi^2}\frac{1}{\epsilon} 
	\left[\lambda_\alpha\left(19\lambda_\alpha + 12\lambda_\Lambda\right)
	+ N \left(2\bar{\xi}\lambda_\alpha + \lambda_m\right)^2
	\right], \\
	\left.iA_{\sigma^3 \Phi} \right\rvert_{\mathrm{div.}} &=
	\frac{3i}{32\pi^2}\frac{1}{\epsilon}\left[12\lambda_\alpha\left(\lambda_\alpha + \lambda_\Lambda\right)
	+ N \left(2\bar{\xi}\lambda_\alpha + \lambda_m\right)\left(\bar{\xi}\lambda_\alpha + \lambda_m\right)
	\right], \\
	\left.iA_{\sigma^2 \Phi^2}\right\rvert_{\mathrm{div.}} &=
	\frac{i}{32\pi^2}\frac{1}{\epsilon}\left[4\lambda_\alpha\left(5\lambda_\alpha + 9\lambda_\Lambda\right)
	+ N \left(2\left(\bar{\xi}\lambda_\alpha + \lambda_m\right)^2 + \lambda_m\left(2\bar{\xi}\lambda_\alpha + \lambda_m\right)\right)
	\right], \\
	\left.iA_{\sigma \Phi^3}\right\rvert_{\mathrm{div.}} &=
	\frac{3i}{32\pi^2}\frac{1}{\epsilon} \left[3\lambda_\alpha\left(\lambda_\alpha + 4\lambda_\Lambda\right)
	+ N \left(\bar{\xi}\lambda_\alpha + \lambda_m\right) \lambda_m
	\right], \\
	\left.iA_{\Phi^4}\right\rvert_{\mathrm{div.}} &=
	\frac{3i}{32\pi^2}\frac{1}{\epsilon} \left[\lambda_\alpha \left(\lambda_\alpha + 12\lambda_\Lambda\right)+ N \lambda_m^2
	\right],
\end{align}
for the four-point functions involving only the conformal mode and the scalaron, and
\begin{align}
	\left.iA_{\sigma^2 \phi_i \phi_j}\right\rvert_{\mathrm{div.}} &=
	\frac{i}{32\pi^2}\frac{\delta_{ij}}{\epsilon} \left[
	6\bar{\xi}\left(2\bar{\xi}+1\right)\lambda_\alpha^2 + \lambda_\alpha \lambda_m \left(1+8\bar{\xi}\right)
	+ 2\left(N+2\right)\left(2\bar{\xi}\lambda_\alpha + \lambda_m\right)\left(\bar{\xi}^2 \lambda_\alpha + \lambda\right)
	\right], \\
	\left.iA_{\sigma \Phi \phi_i \phi_j}\right\rvert_{\mathrm{div.}} &=
	\frac{i}{32\pi^2}\frac{\delta_{ij}}{\epsilon} \left[
	4 \bar{\xi}\lambda_\alpha^2 \left(2\bar{\xi}+1\right) + \left(1+8\bar{\xi}\right)\lambda_\alpha \lambda_m 
	+ 2\left(N+2\right)\left(\bar{\xi}\lambda_\alpha + \lambda_m\right)\left(\bar{\xi}^2 \lambda_\alpha + \lambda\right)
	\right], \\
	\left.iA_{\Phi^2 \phi_i \phi_j}\right\rvert_{\mathrm{div.}} &=
	\frac{i}{32\pi^2}\frac{\delta_{ij}}{\epsilon} \left[
	2\bar{\xi}\left(2\bar{\xi}+1\right)\lambda_\alpha^2 + \left(1+8\bar{\xi}\right)\lambda_\alpha\lambda_m 
	+ 2\left(N+2\right)\lambda_m\left(\bar{\xi}^2 \lambda_\alpha + \lambda\right)
	\right], \\
	\left.iA_{\phi_i \phi_j \phi_k \phi_l }\right\rvert_{\mathrm{div.}} &=
	\frac{i}{16\pi^2}\frac{1}{\epsilon} \left(\delta_{ij}\delta_{kl} + \delta_{ik}\delta_{jl} + \delta_{il}\delta_{jk}\right) \left[
	\bar{\xi}^2 \lambda_\alpha^2 + 2\left(N+8\right)\left(\bar{\xi}^2 \lambda_\alpha + \lambda\right)^2
	\right],
\end{align}
for the four-point functions involving the scalar fields $\phi_i$.

\subsection{Counterterms}

Now we list the divergent parts of the counterterms.
It is important to notice that 
we have to make use of an $\mathrm{SO}(1,1)$ redundancy to renormalize the theory.
We introduce the counterterms to the scalar potential $V$ as
\begin{align}
	V &= \frac{\lambda_\alpha + \delta\lambda_\alpha}{4} 
	\left[ \left(1 +  \theta\right) \left(\sigma + \Phi\right) \left(\sigma +  \theta \Phi\right) + \left(\bar{\xi} + \delta \bar{\xi}\right) \phi_i^2 \right]^2
	+ \frac{\lambda_\Lambda + \delta\lambda_\Lambda}{4}\left(1+4\theta\right)\left(\sigma + \Phi\right)^4 \nonumber \\
	&+ \frac{\lambda_m + \delta \lambda_m}{4}\left(1+2\theta\right)\left(\sigma + \Phi\right)^2\phi_i^2
	+ \frac{\lambda + \delta \lambda}{4}\phi_i^4,
\end{align}
where $\theta$ reflects the $\mathrm{SO}(1,1)$ redundancy that corresponds to the shift
\begin{align}
	\sigma \rightarrow \sigma \cosh \theta + \Phi \sinh \theta,~~
	\Phi \rightarrow \Phi \cosh \theta + \sigma \sinh \theta.
\end{align}
Here we expand the fields with respect to $\theta$ around $\theta = 0$ to first order.
Note that there are only six counterterms, 
$\delta \lambda_\alpha, \theta, \delta \lambda_\Lambda, \delta \lambda_m, \delta \bar{\xi}, \delta \lambda$,
while there are nine divergent four-point functions. 
Hence it is a non-trivial check of our computation that we can indeed renormalize the theory.
After renormalization, the divergent parts of the counterterms are given by
\begin{align}
	\delta \lambda_\alpha &= \frac{1}{32\pi^2}\frac{\lambda_\alpha^2}{\epsilon}\left[5 + 2N\bar{\xi}^2\right], \\
	\theta &=  \frac{1}{32\pi^2}\frac{1}{\epsilon}\left[2\lambda_\alpha + N\bar{\xi} \lambda_m
	\right], \\
	\delta \lambda_\Lambda &=  \frac{1}{64\pi^2}\frac{1}{\epsilon}
	\left[\lambda_\alpha \left(\lambda_\alpha - 4\lambda_\Lambda\right)
	+ N\lambda_m\left(\lambda_m - 8\bar{\xi} \lambda_\Lambda\right)
	\right], \\
	\delta \lambda_m &=  \frac{1}{32\pi^2}\frac{1}{\epsilon}\left[2\bar{\xi}\left(2\bar{\xi}-1\right)\lambda_\alpha^2
	+ \left(-3 + 8\bar{\xi} + 4\bar{\xi}^2\right)\lambda_\alpha \lambda_m - 2N\bar{\xi}\lambda_m^2 
	+ 2\left(N+2\right)\lambda_m \lambda
	\right], \\
	\delta \bar{\xi} &=  \frac{1}{32\pi^2}\frac{1}{\epsilon}\left[\bar{\xi}\lambda_\alpha\left(-3 + 4\bar{\xi} + 4\bar{\xi}^2\right) 
	+ 2\left(N+2\right)\bar{\xi}\lambda
	\right], \\
	\delta \lambda &= \frac{1}{16\pi^2}\frac{1}{\epsilon}\left[\left(2\bar{\xi}-1\right)^2 \bar{\xi}^2 \lambda_\alpha^2 
	+ 12\bar{\xi}^2 \lambda_\alpha \lambda
	+ \left(N+8\right)\lambda^2
	\right].
\end{align}
One can readily derive the one-loop RGEs from these expressions.

\section{Computational details of two-loop RGEs}
\label{app:two_loop}
In this appendix, we explain how we obtain the full two-loop RGEs of the Higgs-$R^2$ theory
shown in Sec.~\ref{subsec:two_loop} with the help of \texttt{SARAH}.
Our main purpose in this appendix is to explain how to take into account the minus sign
in front of the kinetic term of the conformal mode $\Phi$ without modifying the public code itself.
For this purpose, it is important to notice that once we replace 
\begin{align}
	\Phi = i\phi,	
\end{align}
the kinetic term of $\phi$ has the conventional sign.
After this replacement, the coupling between $\phi$ and the other fields 
is multiplied by a power of the imaginary unit $i$.
It suggests that we can take into account the negative sign of the kinetic term of $\Phi$
by putting factors of the imaginary unit $i$ in front of the couplings.
One may feel uncomfortable with the imaginary couplings,
but they should be understood merely as a mathematical trick to translate
an output of public codes/literature (which usually assume the positive sign for the scalar kinetic terms)
to our theory with the ghost-like mode $\Phi$.\footnote{
	At a deeper level, it might be related to the $PT$ symmetry.
}
We again emphasize that, although ghost-like, $\Phi$ does not cause any issue thanks to the residual gauge symmetry.
In the following, we explain how the above idea is implemented in practice.
We divide the procedure into five steps.
Although we focus on the Higgs-$R^2$ theory, our procedure can be equally applied to other models.

\paragraph{Step 1: compute the RGEs of a ``usual" theory.}
As a first step, we compute the RGEs of a ``usual" theory, \textit{i.e.},
the theory that contains two singlet scalar fields $\phi$ and $\sigma$ in addition to the SM,
which are later mapped to the conformal mode of the metric $\Phi$, and the scalaron $\sigma$,
respectively.
We take the scalar potential to be generic, or
\begin{align}
	V = \frac{\lambda_1}{4}\phi^4 + \frac{\lambda_2}{4}\phi^3 \sigma + \frac{\lambda_3}{4}\phi^2 \sigma^2
	+ \frac{\lambda_4}{4}\phi \sigma^3 + \frac{\lambda_5}{4}\sigma^4 
	+ \left(\frac{\kappa_1}{2}\phi^2 + \kappa_2 \phi \sigma + \frac{\kappa_3}{2}\sigma^2\right) \left\lvert H \right\rvert^2
	+ \lambda_\mathrm{UV} \left\lvert H \right\rvert^4,
\end{align}
where $\lambda_i$, $\kappa_i$ and $\lambda_\mathrm{UV}$ are the scalar quartic couplings.
At this moment, both singlets $\phi$ and $\sigma$ are assumed to have positive kinetic terms,
and hence we can easily compute the RGEs of this system with the help of public codes such as \texttt{SARAH},
or literature on the general RGEs such as Refs.~\cite{Machacek:1983tz,Machacek:1983fi,Machacek:1984zw,Luo:2002ti}.

\paragraph{Step 2: map to the Higgs-$R^2$ theory.}
After computing the RGEs of the above theory, we map it to the Higgs-$R^2$ theory.
The scalar potential of the Higgs-$R^2$ theory is given by
\begin{align}
	V &= \lambda \abs{H}^4 + \frac{\lambda_m}{2}e^{2\theta} \left(\Phi + \sigma\right)^2 \abs{H}^2
	+ \frac{\lambda_\Lambda}{4}e^{4\theta}\left(\Phi + \sigma\right)^4 
	+ \frac{\lambda_\alpha}{4}\left[\left(\sigma \cosh\theta + \Phi \sinh\theta\right)
	e^{\theta}\left(\Phi + \sigma\right) + 2\bar{\xi}\abs{H}^2\right]^2,
\end{align}
where we have also included the SO(1,1) redundancy $\theta$.
Note that $\theta$ is unphysical since we can always replace
\begin{align}
	\sigma \cosh\theta + \Phi\sinh\theta \rightarrow \sigma,
	\quad
	\Phi \cosh\theta + \sigma \sinh\theta \rightarrow \Phi,
\end{align}
without affecting the kinetic terms of $\Phi$ and $\sigma$.
The inclusion of this redundancy $\theta$ is nevertheless important for the renormalization,
as we have already seen in App.~\ref{app:one_loop}.
We replace the scalar modes of the Higgs-$R^2$ theory as
\begin{align}
	\Phi \rightarrow i \phi,
	\quad
	\sigma \rightarrow \sigma.
\end{align}
As a result, the two theories are related via
\begin{alignat}{5}
	\lambda_1 &= \frac{\lambda_\alpha}{4}\left(1-e^{2\theta}\right)^2 + e^{4\theta}\lambda_\Lambda,
	\quad
	&&\lambda_2 &&= ie^{2\theta}\left(\lambda_\alpha \left(1-e^{2\theta}\right) - 4 e^{2\theta}\lambda_\Lambda\right),
	\quad
	&\lambda_3 &= \frac{\lambda_\alpha}{2}\left(1-3e^{4\theta}\right) - 6 e^{4\theta} \lambda_\Lambda, 
	\label{eq:mapping1} \\
	\lambda_4 &= ie^{2\theta}\left(\lambda_\alpha\left(1+e^{2\theta}\right) + 4 e^{2\theta}\lambda_\Lambda\right),
	\quad
	&&\lambda_5 &&= \frac{\lambda_\alpha}{4}\left(1+e^{2\theta}\right)^2 + e^{4\theta}\lambda_\Lambda,
	\quad
	&\lambda_\mathrm{UV} &= \lambda + \bar{\xi}^2 \lambda_\alpha, 
	\label{eq:mapping2} \\
	\kappa_1 &= - e^{2\theta}\lambda_m + \left(1 - e^{2\theta}\right)\bar{\xi}\lambda_\alpha,
	\quad
	&&\kappa_2 &&= i e^{2\theta}\left(\lambda_m + \bar{\xi}\lambda_\alpha\right),
	\quad
	&\kappa_3 &= e^{2\theta}\lambda_m + \left(1 + e^{2\theta}\right)\bar{\xi}\lambda_\alpha.
	\label{eq:mapping3}
\end{alignat}
By inserting these mappings, we can compute the RGEs of the Higgs-$R^2$ theory from the RGEs computed in the step~1.

\paragraph{Step 3: consistency check I.}
Before computing the RGEs of the Higgs-$R^2$ theory, we have to perform a consistency check of our procedure.
The ``usual" theory contains nine scalar quartic couplings $\lambda_i$, $\kappa_i$ and $\lambda_\mathrm{UV}$,
while the corresponding Higgs-$R^2$ theory has only six parameter in the scalar potential, 
$\lambda$, $\lambda_m$, $\lambda_\Lambda$, $\lambda_\alpha$, $\bar{\xi}$ and the redundancy $\theta$.
Thus, there are three redundancies among $\lambda_i$, $\kappa_i$ and $\lambda_\mathrm{UV}$ after the mapping,
given by
\begin{align}
	0 &= -2\lambda_1 + i\lambda_2 + i \lambda_4 + 2\lambda_5, \\
	0 &= -\lambda_1 + i\lambda_2 + \lambda_3 - i\lambda_4 - \lambda_5, \\
	0 &= -\kappa_1 + 2i\kappa_2 + \kappa_3.
\end{align}
We have to check that these redundancies are maintained by the beta functions after the mapping, \textit{i.e.},
\begin{align}
	0 &= -2\beta_{\lambda_1} + i\beta_{\lambda_2} + i\beta_{\lambda_4} + 2\beta_{\lambda_5},
	\label{eq:redundancy1} \\
	0 &= -\beta_{\lambda_1} + i\beta_{\lambda_2} + \beta_{\lambda_3} - i\beta_{\lambda_4} - \beta_{\lambda_5},
	\label{eq:redundancy2} \\
	0 &= -\beta_{\kappa_1} + 2i \beta_{\kappa_{2}} + \beta_{\kappa_3},
	\label{eq:redundancy3}
\end{align}
after replacing the couplings following Eqs.~\eqref{eq:mapping1}-\eqref{eq:mapping3}.
These conditions, once satisfied, guarantee that the Higgs-$R^2$ theory is renormalizable.
We have explicitly checked that these conditions are indeed satisfied up to two-loop
in the Higgs-$R^2$ theory.

\paragraph{Step 4: compute the RGEs of the Higgs-$R^2$ theory.}
Once the above consistency check is done, we are ready to compute the RGEs of the Higgs-$R^2$ theory.
From the mappings~\eqref{eq:mapping1}-\eqref{eq:mapping3}, 
we can obtain the RGEs as
\begin{align}
	\beta_{\lambda_\alpha} &= \frac{1}{2}\left(3\beta_{\lambda_1} + \beta_{\lambda_3} + 3\beta_{\lambda_5}\right), \\
	\beta_{\theta} &= 
	\frac{e^{2\theta}\beta_{\lambda_\alpha}}{2\left(\lambda_1 - \lambda_5\right)}
	+ \frac{\beta_{\lambda_1}-\beta_{\lambda_5}}{2\left(\lambda_1 - \lambda_5\right)}, \\
	\beta_{\lambda_\Lambda} &= -e^{-4\theta}\beta_{\theta}\left(2\lambda_1 + 2\lambda_5-\lambda_\alpha\right)
	+ \frac{e^{-4\theta}}{2}\left(\beta_{\lambda_1}+\beta_{\lambda_5}\right) - \frac{\beta_{\lambda_\alpha}}{4}\left(1+e^{-4\theta}\right), \\
	\beta_{\bar{\xi}} &= 
	\frac{\left(\beta_{\kappa_1}+\beta_{\kappa_3}\right)}{2\lambda_\alpha}
	-\frac{\bar{\xi}}{\lambda_\alpha}\beta_{\lambda_\alpha}, \\
	\beta_{\lambda_m} &=
	-\frac{1}{2}\left(\beta_{\kappa_1} + \beta_{\kappa_3}\right)
	- ie^{-2\theta}\left(\beta_{\kappa_2} -2\kappa_2 \beta_{\theta}\right), \\
	\beta_{\lambda} &=
	\beta_{\lambda_\mathrm{UV}} - 2\bar{\xi}\lambda_\alpha \beta_{\bar{\xi}} - \bar{\xi}^2\beta_{\lambda_\alpha}.
\end{align}
It is convenient to compute the beta functions in this ordering,
as the beta functions given in the later equations are expressed by those computed earlier.
The RGEs of the other SM parameters, such as the Yukawa and gauge couplings,
are also obtained by replacing the parameters following 
the mappings~\eqref{eq:mapping1}-\eqref{eq:mapping3}.

\paragraph{Step 5: consistency check I\hspace{-.1em}I.}
Finally we have two additional consistency checks of our procedure.
First, as we have explained above, the parameter $\theta$ is unphysical 
and hence the beta functions should not depend on it.
Second,
the beta functions $\beta_{\lambda_{\alpha}}$, $\beta_{\bar{\xi}}$ and $\beta_{\lambda}$
should be independent of $\lambda_m$ and $\lambda_\Lambda$,
and $\beta_{\lambda_m}$ should be independent of $\lambda_\Lambda$.
This property is easily understood in the Jordan frame from the mass dimension of the couplings.
It is, however, a non-trivial check that it is satisfied in our formalism since all the parameters are 
treated as scalar quartic couplings once we map the Higgs-$R^2$ theory to the LSM~\eqref{eq:LSM}.
We have checked that these properties are indeed satisfied by the RGEs computed by our procedure.

\vspace{5mm}

The two-loop RGEs in Sec.~\ref{subsec:two_loop} are derived following the above steps~1-5.
We have also checked that the one-loop RGEs computed by this procedure 
agree with those derived from the direct computation in App.~\ref{app:one_loop}.

\section{Renormalizability and residual gauge symmetry}
\label{sec:renormalizability}

In this paper, we have confirmed by explicit computation that the LSM~\eqref{eq:LSM}
is renormalizable (at least) up to two-loop.
In other words, all the divergences that appear in the theory can be cancelled by the operators within the LSM
and hence one does not have to introduce additional operators as counter terms.

We emphasize that the renormalizability of the LSM~\eqref{eq:LSM} 
is non-trivial if we regard $\Phi$ and $\sigma$ as usual scalar fields.
This is because the scalar potential of the LSM~\eqref{eq:LSM}
does not exhaust all the possible scalar quartic interactions.
If one regards $\Phi$ and $\sigma$ as usual scalar fields,
one would expect that there are nine operators,
$\Phi^4$, $\Phi^3 \sigma$, $\Phi^2 \sigma^2$, $\Phi \sigma^3$, $\sigma^4$,
$\Phi^2 \phi_i^2$, $\Phi \sigma \phi_i^2$, $\sigma^2 \phi_i^2$, and $\phi_i^4$,
with independent coefficients.
Instead, the coefficients of these operators are related in the LSM~\eqref{eq:LSM} 
such that there are only six independent parameters (including the $\mathrm{SO}(1,1)$ redundancy).
It is nevertheless renormalizable thanks to the
the ghost-like property of $\Phi$,
as one can see by explicit computation.
In this appendix, we describe our attempt to understand this renormalizability
from the residual gauge symmetry.
In particular, we see that the residual gauge symmetry
restricts possible forms of the scalar quartic interactions
of the LSM~\eqref{eq:LSM}.

\subsection{Residual gauge symmetry and conformal weight}

We first explain the residual gauge symmetry of our gauge fixing condition~\eqref{eq:gauge_fix}.
Here we focus on the scalar part of the residual gauge symmetry,
which is given by
\begin{align}
	x^{\mu} \rightarrow x^{\mu} - \partial^\mu \xi,
\end{align}
where the index is raised by $\eta^{\mu\nu}$, and $\xi$ satisfies
\begin{align}
	\partial_{\mu}\partial_{\nu} \xi = \frac{1}{4}\eta_{\mu\nu} \Box \xi.
\end{align}
It follows that $\xi_\mu \equiv \partial_\mu \xi$ satisfies the conformal killing equation (of flat spacetime),
\begin{align}
	\partial_\mu \xi_\nu + \partial_\nu \xi_\mu = \frac{2}{d}\eta_{\mu\nu} \partial_\alpha \xi^\alpha,
\end{align}
where $d = 4$ is the spacetime dimension,
and hence this residual gauge symmetry is (one form of) the conformal symmetry,
or the dilatation symmetry.
Here one should not confuse the conformal transformation with the Weyl transformation.
The former is a general coordinate transformation under which 
a given metric ($\eta_{\mu\nu}$ in our case) does not transform up to the overall normalization,\footnote{
	Or equivalently, a general coordinate transformation times a Weyl transformation
	that keeps a given metric unchanged.
}
while the latter is a field redefinition of the metric and is unrelated to the general coordinate transformation.\footnote{
	In the literature, the word ``conformal transformation" is sometimes used instead of the Weyl transformation
	in the latter meaning.
	In this paper we follow the terminology we have defined above in order to avoid possible confusion.
}
Note that the above conformal killing equation in particular means that
\begin{align}
	\partial_\mu \Box \xi = \Box^2 \xi = 0.
\end{align}

For our purpose, it is useful to define the conformal weight.
We assign the conformal weight $n_\alpha$ to an operator $O_\alpha$ if it transforms as
\begin{align}
	O_\alpha \rightarrow 
	\left[1 + \frac{n_\alpha}{4}\Box \xi + \left(\partial^\mu \xi\right)\partial_\mu\right]O_\alpha,
\end{align}
under the residual gauge symmetry.
Note that the conformal weight is additive, that is, 
if operators $O_\alpha$ and $O_\beta$ have conformal weights $n_\alpha$ and $n_\beta$, respetively,
the composite operator $O_\alpha O_\beta$ transforms as
\begin{align}
	O_\alpha O_\beta \rightarrow 
	\left[1 + \frac{n_\alpha+n_\beta}{4}\Box \xi + \left(\partial^\mu \xi\right)\partial_\mu\right]O_\alpha O_\beta.
\end{align}
The derivative $\partial_\mu$ raises the conformal weight of a given operator by unity, or
\begin{align}
	\partial_\mu O_\alpha  \rightarrow 
	\left[1 + \frac{n_\alpha+1}{4}\Box \xi + \left(\partial^\nu \xi\right)\partial_\nu \right] \partial_\mu O_\alpha,
\end{align}
as one can show by using the conformal killing equation.
If an operator $O$ has conformal weight four, one can show that
\begin{align}
	\int d^4 x O \rightarrow
	\int d^4 x \left[1 + \Box \xi + \left(\partial^\nu \xi\right)\partial_\nu \right] O
	= \int d^4 x O,
\end{align}
\textit{i.e.}, its integrand is invariant under the residual gauge symmetry.
It follows that only operators with conformal weight four can show up in the Lagrangian
due to the requirement of invariance under the residual gauge transformation.

\subsection{Possible interactions}
Now we write down possible interactions 
of the LSM that are invariant under the residual gauge symmetry.
We start with only $\Phi$ and the SM particles as the particle content.
In other words, we do not include the scalaron as a fundamental degree of freedom at the beginning.

The conformal mode of the metric transforms under the residual gauge symmetry as
\begin{align}
	\Phi \rightarrow \left[1 + \frac{1}{4}\Box \xi + \left(\partial^\mu \xi\right)\partial_\mu\right]\Phi,
\end{align}
and hence it has conformal weight one.
Since we rescaled the SM fields by an appropriate factor of  $e^{\varphi}$,
the SM fields (in the ``comoving frame") transform under the residual gauge symmetry as
\begin{align}
	H &\rightarrow \left[1 + \frac{1}{4}\Box \xi + \left(\partial^\mu \xi\right)\partial_\mu\right]H,  \\
	\psi &\rightarrow \left[1 + \frac{3}{8}\Box \xi + \left(\partial^\mu \xi\right)\partial_\mu\right]\psi,  \\
	A_\nu &\rightarrow \left[1 + \frac{1}{4}\Box \xi + \left(\partial^\mu \xi\right)\partial_\mu\right]A_\nu,
\end{align}
where $H$ is the Higgs doublet, $\psi$ represents the SM fermions and $A_\nu$ the SM gauge bosons.

Since we start from the theory which \textit{does not have the scalaron $\sigma$ explicitly}, 
the operators with conformal weight four that are written solely in terms of $\Phi$ are
\begin{align}
	\left(\partial \Phi\right)^2,
	\quad
	\left(\frac{\Box \Phi}{\Phi}\right)^2,
	\quad
	\Phi^4.
	\label{eq:pure_gravity}
\end{align}
They correspond to the operators $R$, $R^2$ and the cosmological constant, respectively.
Of course one can write down more terms by using more derivatives, 
but they are higher-dimensional operators and
hence we ignore them below.
In the purely SM sector,
all the usual SM interactions are allowed.
For the SM-$\Phi$ mixed sector, the following are the leading order terms that
respect both the residual gauge symmetry and the SM gauge symmetry:
\begin{align}
	\left\lvert H \right\rvert^2 \Phi^2,
	\quad
	\left\lvert H \right\rvert^2
	\frac{\Box \Phi}{\Phi},
	\label{eq:SM_gravity}
\end{align}
where the first one corresponds to the Higgs mass term and the second one to the non-minimal coupling to gravity $\xi$
(or more precisely $\bar{\xi}$).
Again we can write down other operators but we omit them due to their higher-dimensional nature.
It follows that the scalar quartic interactions of the LSM~\eqref{eq:LSM}
are determined by the residual gauge symmetry.

\subsection{A few remarks}
We have seen above that the structure of the theory and hence the scalar potential 
are controlled by the residual gauge symmetry. 
We still feel that there are unsatisfactory points in this argument.
In this subsection we point out some of them
in order to motivate possible future work on the theoretical structure of the Higgs-$R^2$ theory.

The first unsatisfactory point is that we had to assume that the scalaron is not a fundamental degree of freedom
at the beginning in our discussion above.
The reason is that if we include the scalaron $\sigma$ from the beginning,
the scalar part of the residual gauge symmetry does not distinguish $\sigma$ from usual singlet scalar fields,
and hence it does not prohibit generic scalar quartic interactions involving $\sigma$.
Moreover, we can still write down $(\Box \Phi/\Phi)^2$ which would even introduce an additional ``scalaron."
We think that there should be a propertry of $\sigma$ that distinguishes it from other usual scalar fields,
whose study we leave as a future work.

The second unsatisfactory point is the criteria of higher dimensional operators.
In the discussion above, we have included only the SM operators and the operators in Eqs.~\eqref{eq:pure_gravity}
and~\eqref{eq:SM_gravity} by arguing that others are higher dimensional operators.
A rule of thumb is that we count the mass dimensions of the SM particles and the derivative as usual,
but we count that of $\Phi$ as zero.
We also allow only $\Phi$ to appear in the denominator of operators.
A given operator is then regarded as higher dimensional if the total mass dimensions exceeds four.
We know that this criteria practically works, but we do not know a physical reason behind it.
We again leave this point for future work.

\small
\bibliographystyle{utphys}
\bibliography{ref}

\end{document}